\documentclass[aps,twocolumn,reprint]{revtex4-1}
\usepackage{graphics}
\usepackage{caption}
\usepackage{subcaption}
\usepackage{epstopdf}
\usepackage{epsfig}
\usepackage{epsf,epic}
\usepackage{color}
\usepackage{marvosym }
\usepackage{siunitx}
\usepackage{amsmath}
\usepackage{amssymb}
\usepackage{amsfonts}
\usepackage{wrapfig}
\usepackage{pstricks}
\usepackage{multirow}
\usepackage{pst-node}
\usepackage{extdash}
\usepackage{braket}
\usepackage{bm}
\usepackage{dcolumn}% Align table columns on decimal point
\newcommand{\etal}{\textit{et al\ }}

\begin{document}
\title{Quasiparticle self-consistent $GW$ band structures and high-pressure phase transitions of LiGaO$_2$ and NaGaO$_2$}
\author{Santosh Kumar Radha, Amol Ratnaparkhe  and Walter R. L. Lambrecht}
\affiliation{Department of Physics, Case Western Reserve University, 10900 Euclid Avenue, Cleveland, Ohio 44106-7079, USA}
\begin{abstract}
  Quasi-particle self-consistent $GW$ calculations are presented for the
  band structures of LiGaO$_2$ and NaGaO$_2$ in the orthorhombic $Pna2_1$
  tetrahedrally coordinated crystal structures, which are closely
  related to the wurtzite structure of ZnO. Symmetry labeling of the bands near the gap is carried out and effective mass tensors  are extracted for the
  conduction band minimum and crystal field split valence band maxima at $\Gamma$. The gap is found to be direct at $\Gamma$ and is 5.81 eV in LiGaO$_2$ and
  5.46 eV in NaGaO$_2$.  Electron-phonon coupling zero-point normalization
  is estimated to lower these gaps by about 0.2$\pm0.1$  eV.
  Optical response functions are calculated within the independent particle
  long wavelength limit and show the expected anisotropy of the absorption
  onsets due to the crystal field splitting of the VBM. The results 
  show that both materials are
  promising candidates as ultrawide gap semiconductors with wurtzite
  based tetrahedrally bonded crystal structures. Direct transitions
  from the lowest conduction band to higher bands, relevant to n-type
  doped material and transparent conduction applications are found
  to start only above 3.9 eV and are allowed for only one polarization,
  and several higher band transitions are forbidden by symmetry. 
  Alternative crystal structures, such as
  $R\bar{3}m$ and a rocksalt type phase with tetragonally distorted $P4/mmm$
  spacegroup, both 
  with octahedral coordination of the cations are also investigated.
  They are found to have higher energy but about 20 \% smaller volume
  per formula unit.
  The transition pressures to these phases are determined and for LiGaO$_2$
  found to be in good agreement with experimental studies. The $R\bar{3}m$
  phase also has a comparably high but slightly indirect band gap
  while the rocksalt type phase if found to have a considerably smaller
  gap of about 3.1 eV in LiGaO$_2$ and 1.0 eV in NaGaO$_2$.  
\end{abstract}
\maketitle
\section{Introduction}
LiGaO$_2$ is a transparent ceramic material which has been
considered for piezoelectric\cite{Nanamatsu72,Gupta76,Boonchun10}
and non-linear optical applications \cite{Rashkeev99} in the past
and can be grown in bulk single crystal form,\cite{Marezio65,Ishii98}
which has among other led to its use as closely lattice matched
substrate\cite{Sakurada07} for GaN epitaxial growth.  
It can be viewed as a I-III-VI$_2$ analog of the II-VI material
ZnO with a wurtzite based crystal structure, consisting of an
ordered arrangement of the Li and Ga atoms on the cation sublattice
of the wurtzite. In particular it has the $Pna2_1$ spacegroup.
Mixed alloy systems of ZnO and LiGaO$_2$  and 
ZnO/LiGaO$_2$ heterojunctions have
also been studied.\cite{Omata08,Omata11,Ohkubo2002}

Although mostly considered an insulating material, it has recently been
proposed that LiGaO$_2$ can be doped n-type with Si or Ge, which would make
it promising for ultrawide gap
semiconductor applications.\cite{Boonchun11SPIE,Boonchun19,Boonchun20,Lenyk18,Skachkov20}
From this point of view it may have some advantages relative to the now 
widely pursued $\beta$-Ga$_2$O$_3$.\cite{Sasaki13,Green16}
It has a simpler crystal structure
with all atoms tetrahedrally coordinated and it appears to have an
even wider band gap.

However, the band gap is not yet fully established.  While experiments
indicate a gap of about 5.3-5.6 eV, a prior $GW$ calculations predicted
an even larger gap of 6.25 eV.\cite{Boonchun11SPIE}
Optical absorption data give a direct gap of 
5.5 eV \cite{Wolan98,Ohkubo2002} to 5.26 eV \cite{Chen14} 
while X-ray absorption and emission data \cite{Johnson2011} gives
a gap of 5.6 eV. Boonchun and Lambrecht\cite{Boonchun11SPIE}
tried to explain their band gap overestimate compared to experiment
in terms of  temperature dependence of the gaps and zero-point motion
correction by electron-phonon coupling. However, at that time no
accurate predictions of these effects were possible.
Since then, we have found in various other systems that
{\bf k}-point and basis set
convergence can significantly affect the QS$GW$ band gap
results.\cite{Amolalgo,Ratnaparkhe17}
In this paper we re-evaluate the band structure of LiGaO$_2$
with well-converged quasiparticle self-consistent $GW$ calculations
and review the estimates of the electron-phonon coupling effects.

Secondly, we consider a related material of the same family,
NaGaO$_2$ to evaluate the possibility of band gap tuning by
varying the alkali metal component. One of the practical problems
found in the past with LiGaO$_2$ as substrate is the ionic mobility of Li,
which tends to easily diffuse.  In particular for high-power applications with
at high temperature or in the presence of strong electric fields, ion mobility
might be expected to be a problem. Therefore replacing it by a less diffusive
element Na might be beneficial. Indeed one of the main attractive features
of ultrawide gap semiconductors is their large breakdown field.
But then we also need to ensure that these high fields do not lead to
ionic diffusion or loss of Li from the system.

The known ground state structure of LiGaO$_2$ is the $Pna2_1$ structure,\cite{Marezio65} for which the prototype is $\beta$-NaFeO$_2$.
In this standard setting of the space group, 
$b>a>c$ with $b\approx2a_w$, $a\approx\sqrt{3}a_w$ and $c=c_w$
in relation to the wurtzite hexagonal lattice constants. 
Note that in some previous literature,\cite{Boonchun10}
the lattice constants $a$ and $b$ are reversed, 
$a>b>c$ in which case the space group setting is $Pbn2_1$. 
The notation $n$ corresponds to a glide mirror plane with two glides,
$a$ or $b$ to a glide plane with one glide along the indicated direction,
so in the $Pna2_1$ notation $n$ is in the $bc$ plane and $a$ in the $ac$
plane and in both cases $2_1$ indicates a two-fold screw-axis along $c$.

However, it is
also worthwhile to study the competing $\alpha$-NaFeO$_2$ structure, which
has spacegroup $R\bar{3}m$. In that structure, all atoms are octahedrally
coordinated and the structure can be viewed as a layered structure.
According to Materials Project (MP)\cite{MP} this structure has higher energy
than the $Pna2_1$ structure for LiGaO$_2$ by 61 meV/atom. 
This structure is commonly found in many ABO$_2$ systems with metallic elements
functioning as cations. For example this structure is found for
LiCoO$_2$, a well known battery material, in  which the Li content
can to some extent be varied by chemical or electrochemical means.
Since NaGaO$_2$ has not been reported to the best of our knowledge,
it is important to check the relative stability of the two structures
and determine the energetic preference for octahedral {\sl vs.} tetrahedral
coordination of the Ga and alkali elements Li and Na. We therefore here
also study the $R\bar{3}m$ band structures and relative stability of
the two structures. In LiGaO$_2$, disordered rocksalt phases have also
been reported and these are also considered here. 

Finally, besides the electronic band structure, which is here provided
in more detail than in Ref. \onlinecite{Boonchun11SPIE}, we also
study the interband transition response functions related to optical
absorption. 

\section{Computational methods}
The calculations in this work are done using the full-potential linearized
muffin-tin orbital (FP-LMTO) all-electron method within either
density functional theory (DFT) or many-body-perturbation theory (MBPT)
context. The FP-LMTO method is used as implemented in the questaal
package,\cite{questaal,questaalpaper} and based on the work by
Methfessel \etal\cite{Methfessel} and since then improved to allow
for inclusion of augmented plane waves as additional basis
functions,\cite{Kotani10} which allows for a systematic check of
the basis set convergence. We start our calculations from the
structures available at the Materials Project (MP) \cite{MP}
and then check the smallness of the residual forces within FP-LMTO
or further relax the atomic positions. For materials not yet available
in MP, we used the Quantum Espresso code \cite{Giannozzi_2009}
to relax the atomic positions and
lattice constants simultaneously before additional testing with
FP-LMTO.As DFT functional, we use the Perdew-Burke-Ernzerhof (PBE)\cite{PBE}
generalized gradient approximation (GGA).

A main advantage of the
questaal package is that it has one of the few all-electron implementations
of the $GW$ MBPT method. Here $G$ and $W$ refer to the one-particle
Green's function and screened Coulomb interaction $W$,\cite{Hedin65,Hedin69}
which define the self-energy operator $\Sigma=iGW$ in a schematic notation. 
Furthermore, this implementation uses a
mixed-product-interstitial-plane-wave basis instead of only plane waves
to represent all two-point quantities, such as the bare Coulomb interaction $v$,
screened Coulomb interaction $W=\varepsilon^{-1}v=[1-vP]^{-1}v$,
polarization propagator $P$ and inverse dielectric
response function $\varepsilon^{-1}$. This representation is far 
more efficient to represent the response and does not require one
to include as many high-energy empty bands for convergence.
Details of the $GW$ implementation can be found in
Kotani \etal\cite{Kotani07} and Ref. \onlinecite{questaalpaper}.

The $GW$ method is here used in the quasiparticle self-consistent version,
known as QS$GW$.  In this approach, the energy dependent $\Sigma(\omega)_{ij}$
is replaced by an energy-independent Hermitian average
$\tilde\Sigma_{ij}=\frac{1}{2}\mathrm{Re}[\Sigma_{ij}(\epsilon_i)+\Sigma_{ij}(\epsilon_j)]$, 
represented in the basis  of initial $H^0$ eigenstates, where
$H^0$ is the DFT starting Hamiltonian. The $\tilde{\Sigma}_{ij}-v^{DFT}_{xc}$
is then added to the $H^0$ Hamiltonian in each iteration, providing
a new $G^0$ Green's function from which a new $W^0$ and $\Sigma^0=iG^0W^0$
is obtained in the next step. At convergence, the eigenvalues of the
Kohn-Sham Hamiltonian $H^0$  are equal to the quasiparticle energies.
Hence the name {\sl quasiparticle self-consistent}. In other words,
we are here focused on obtaining the real
{\sl quasiparticle  energies}, independent
of the DFT starting point, rather than the full energy dependent
complex self-energy or Green's function which would contain a more comprehensive
description of the quasiparticle spectral function.

Thanks to the atom-centered LMTO basis set, to which the self-energy
can be converted, a natural route to interpolating the self-energy
eigenvalue shifts to other {\bf k}-points than the mesh on which
$\tilde\Sigma({\bf k})_{ij}$ is calculated is available. Hence $GW$-accuracy
energy bands and effective masses 
are obtained along the symmetry lines, or, on a fine
mesh for density of states or optical response functions,
without the need for the computationally expensive 
evaluation of $\tilde\Sigma({\bf k})_{ij}$ on an equally fine mesh.
Nonetheless, the {\bf k}-mesh
on which the $GW$ self-energy is determined is important for convergence.
One finds that  a coarse mesh tends to give larger band
gaps.\cite{Kotani07,vanSchilfgaarde06,Amolalgo} Also important
are a large  basis set including typically $spdf-spd$ angular momentum
channels for two sets of smoothed Hankel function envelopes of the
LMTOs as well as additional local orbitals to represent either  semi-core
states or higher lying conduction band contributions
to the partial waves of the same angular momentum
character within the muffin-tin sphere partial.

We here used  $3\times3\times3$ and $4\times4\times4$ {\bf k}-meshes
on which the $\tilde\Sigma$ is calculated
for the $Pna2_1$ structure to check convergence and a $6\times6\times6$
mesh for the $R\bar{3}m$ and $P4/mmm$ structures. 

\section{Results}
\subsection{Structural properties and stability of LiGaO$_2$}
\begin{table*}
  \caption{Wyckoff positions and symmetry operation
    linking equivalent sites, lattice constants, volume per formula unit,
    reduced coordinates and bond lengths for LiGaO$_2$
    in $Pna2_1$ structure, comparing expt. data from Ref. \onlinecite{Marezio65} with PBE-GGA relaxed structure.\label{tabstruc}}
  \begin{ruledtabular}
    \begin{tabular}{lcccc}
      4a positions   &$x,y,z$  & $-x,-y,z+\frac{1}{2}$ & $\frac{1}{2}-x,y+\frac{1}{2},z+\frac{1}{2}$ & $x+\frac{1}{2}, \frac{1}{2}-y,z$ \\
      operation & $1$ & $2_{1z}$ & $n_x$ & $a_y$ \\  \hline
      \multicolumn{5}{c}{Expt. \cite{Marezio65}} \\ \hline
      $a$ (\AA)  & $b$ (\AA) & $c$ (\AA) & $V/fu$ (\AA$^3$) \\ \hline
      5.407      & 6.405 & 5.021 & 43.471 \\ 
             $2a/b$ & $2c/b$ & $b/2$ \\
      1.6884 & 1.5678 & 3.2025  \\ \hline
      atom & Wyckoff & $x$ & $y$ & $z$ \\ \hline 
      Li   & 4a      & 0.0793 & 0.6267 & $-$0.0064 \\
      Ga   & 4a      & 0.0821 & 0.1263 & 0.0000 \\
      O$_\mathrm{Li}$ & 4a &0.0934 & 0.6388 & 0.3927 \\
      O$_\mathrm{Ga}$ & 4a & 0.0697 & 0.1121 & 0.3708 \\ \hline
      &\multicolumn{4}{c}{bond lengths (\AA)} \\ \hline
       Ga-O$_\mathrm{Ga}^c$ & Ga-O$_\mathrm{Ga}^a$ & Ga-O$_\mathrm{Li}^a$ & Ga-O$_\mathrm{Li}^b$ \\ 
       1.865              & 1.851 & 1.837 & 1.858 \\
       Li-O$_\mathrm{Li}^c$ & Li-O$_\mathrm{Li}^a$ & Li-O$_\mathrm{Ga}^a$ & Li-O$_\mathrm{Ga}^b$ \\
       2.007               & 2.005               & 1.998               & 1.957 \\ \hline\hline
       \multicolumn{5}{c}{GGA-PBE\cite{MP}} \\ \hline
      $a$ (\AA)  & $b$ (\AA) & $c$ (\AA) & $V/fu$ (\AA$^3$) \\ \hline
      5.4665      & 6.4570 & 5.094 & 44.952 \\ 
             $2a/b$ & $2c/b$ & $b/2$ \\
      1.6932 & 1.5778 & 3.228  \\ \hline
      atom & Wyckoff & $x$ & $y$ & $z$ \\ \hline 
      Li   & 4a      & 0.0823 & 0.6244 & 0.0001 \\
      Ga   & 4a      & 0.0811 & 0.1261 & 0.0046 \\
      O$_\mathrm{Li}$ & 4a & 0.0928 & 0.6382 & 0.3959\\
      O$_\mathrm{Ga}$ & 4a & 0.0688 & 0.1125 & 0.3726 \\ \hline
      &\multicolumn{4}{c}{bond lenths (\AA)} \\ \hline
       Ga-O$_\mathrm{Ga}^c$ & Ga-O$_\mathrm{Ga}^a$ & Ga-O$_\mathrm{Li}^a$ & Ga-O$_\mathrm{Li}^b$ \\ 
       1.878              & 1.877 & 1.868 & 1.870 \\
       Li-O$_\mathrm{Li}^c$ & Li-O$_\mathrm{Li}^a$ & Li-O$_\mathrm{Ga}^a$ & Li-O$_\mathrm{Ga}^b$ \\
       2.019               & 2.018               & 2.016               & 1.998 \\
    \end{tabular}
  \end{ruledtabular}
\end{table*}

We start by examining the calculated and experimental structural parameters
of LiGaO$_2$ in
the $Pna2_1$ structure in Table \ref{tabstruc}.
This table also shows  which symmetry operations link the different equivalent
atoms of the $4a$ Wyckoff position. Here, $2_{1z}$ is the twofold screw axis
along $z$ located at the origin. The $n_x$ is a double glide plane perpendicular
to $x$ and with glides by $b/2$ and $c/2$ which occurs at $x=\frac{1}{4}$
and therefore also involves a shift by $a/2$. The $a_y$ is a single-glide
plane perpendicular to $y$ with shift by $a/2$ but it occurs at $y=1/4$
and hence also involves a shift by $b/2$. Here the Cartesian axes
$x,y,z$ are chosen along $a,b,c$ respectively.

We here compare the experimental structural
parameters with the calculated ones within the PBE-GGA density functional
from Materials Project.\cite{MP} The reduced coordinates within that model
were verified using the FP-LMTO method and agree to $\pm0.001$.
We can see that PBE overestimates each of the lattice constants
by about 1 \% and hence the volume by 3 \%. Interestingly, it
overestimates Ga-O bond lenths slightly more than Li-O bond lengths.
Also it overestimates $c$ aby about 1.4 \% and $a$ and $b$ by about 0.8 \%.
Each cation has four different bond lengths to oxygen, for example
Ga-O$_\mathrm{Li}^a$ means the bond length between Ga and the O$_\mathrm{Li}$
type O  in the $a$-direction. We can see that the Li-O bond lengths
are significantly larger than the Ga-O bond lengths.
The $2a/b$ ratio in the undistorted wurtzite structure derived
$Pna2_1$ structure would be $\sqrt{3}\approx1.732$, but here is
reduced to 1.69. This implies that the 120$^\circ$ angle between two wurtzite
lattice vectors in the plane has here increased to 122$^\circ$.

\begin{table}
  \caption{Structural parameters of LiGaO$_2$ in $R\bar{3}m$ structure. \label{tabstrucr-3m}}
  \begin{ruledtabular}
    \begin{tabular}{cccc}
       & Li & Ga & O \\ \hline
      Wyckoff &  1a & 1b & 2c \\ 
      reduc. coord. & $(0,0,0)$ & $(\frac{1}{2},\frac{1}{2},\frac{1}{2})$ & $(\pm u,\pm u,\pm u)$ \\ \\ \hline
      \multicolumn{4}{c}{Calculated} \\ \hline 
      $a=b=c$ (\AA) & $\alpha=\beta=\gamma$ & $V/fu$ (\AA$^3$) & $u$ \\ \hline
      5.173 & 33.138$^\circ$                & 36.824 & 0.2415 \\ \hline
      bond lengths (\AA) & Li-O & Ga-O \\ \hline 
      & 2.171 & 2.026 \\ \hline
      \multicolumn{4}{c}{Experiment\cite{MarezioRemeika}} \\ \hline
        $a=b=c$ (\AA) & $\alpha=\beta=\gamma$ & $V/fu$ (\AA$^3$) & $u$ \\ \hline
        5.1066$\pm0.0005$ & 33.12$^\circ$ & 35.394 & 0.2417 \\
        bond lengths (\AA) & Li-O & Ga-O \\ \hline
         & 2.14 & 2.00 \\ 
    \end{tabular}
  \end{ruledtabular}
\end{table}

Next, we discuss the structural stability relative to the $R\bar{3}m$
structure. Note that in the $R\bar{3}m$ structure, the cations
have octahedral coordination and form a layered structure with alternating
Li and Ga containing layers. We start again from the structural parameters
of the Materials Project,\cite{MP}
which are optimized within the GGA-PBE density functional.
The structural parameters of $R\bar{3}m$  are given in
Table \ref{tabstrucr-3m}.
Clearly, the volume per formula unit
in this structure is significantly (18 \%) lower than in the $Pna2_1$
structure, meaning that this phase can be stabilized under pressure.
Interestingly, both Ga-O and Li-O bond lengths are larger in the
octahedral environment. 
This structure is known as $\alpha$-LiGaO$_2$ and its structure was determined by Marezio and Remeika \cite{MarezioRemeika}.

Yet, another form of LiGaO$_2$ is known as $\gamma$-LiGaO$_2$ and has a
rocksalt-like structure with tetragonal distortion.  In this phase, LiGaO$_2$
is also octahedrally coordinated and the Li and Ga occur in a disordered
way in the Wyckoff $2b$ positions of space group $I4/m$ with equal
probability. Instead, a closely related structure, $P4/mmm$ is considered
in MP\cite{MP}. This is an ordered distorted rocksalt modification.
It consist of alternating layers of Li and Ga in the rocksalt structure
along one of the cubic axes, say the [001] axis. This axis now becomes
the $c$-axis after distortion.  The lattice constants and atomic
positions of this phase from MP\cite{MP} are given in
Table \ref{tabstrucp4mmm}. Note that in this structure the
Li-O and Ga-O bond lengths are fixed to be $a\sqrt{2}/2$ perpendicular
to $c$ and $c/2$ parallel to $c$. This is unfavorable from the point of
view that the bond lengths cannot be individually optimized for
each species.  The experimental $I4/m$ disordered rocksalt-like
structure according to Lei \etal \cite{Lei10}
has lattice constants of $a=2.8763$ \AA\ and $c=4.1929$ \AA\
and hence $c/a=1.4577$, close to the ideal $\sqrt{2}=1.4142$. In contrast,
the $c/a$ ratio in $P4/mmm$ is reduced to 1.286.
Following Lei \etal \cite{Lei10} a disordered cubic rocksalt
phase with space group $Fm\bar{3}m$ also exists and is designated as
$\delta$-LiGaO$_2$. It has a lattice constant of 4.1134 \AA. All
these phases have close volumes per formula unit of about 34.8$\pm0.1$  \AA$^3$.

\begin{table}
  \caption{Lattice constants and atomic positions of LiGaO$_2$ in $P4/mmm$ structure.\label{tabstrucp4mmm}}
  \begin{ruledtabular}
    \begin{tabular}{lcccc}
      atom & Wyckoff  & $x$ & $y$ & $z$  \\ \hline
      Li  & 1c       & 0.5    & 0.5     & 0.0  \\
      Ga  & 1b       & 0.0    & 0.5     & 0.0  \\
      O   & 1a       & 0.0    & 0.0     & 0.0 \\
      O   & 1d       & 0.5    & 0.5     & 0.5 \\ \hline
  lattice constants (\AA)    & $a=b$        & $c$ & $V/fu$  \\ \hline
     &  3.002       & 3.861  & 34.807 \\
  bond lengths (\AA)   & $\parallel c$ & $\perp c$ \\
   &   1.930 & 2.123 \\
    \end{tabular}
\end{ruledtabular}   
\end{table}

\begin{table}
  \caption{Cohesive energy ($E_0/f.u.$), equilibrium volume ($V_0$), bulk modulus ($B_0$)
    and its pressure derivative ($B_0^\prime$) and transition pressure, $p_t$
    from $Pna2_1$ to other phase for LiGaO$_2$. \label{tabeos}}
  \begin{ruledtabular}
    \begin{tabular}{lccc}
    property  & $R\bar{3}m$ &  $Pna2_1$ & $P4/mmm$ \\ \hline
    $V_0$ (\AA$^3$)    & 36.297 & 44.424 & 34.712 \\
    $E_0$ (eV/f.u.)    & 21.91 & 22.30 & 21.62\\
    $B_0$ (GPa)        & 166 & 130 & 1615 \\
    $B_0^\prime$        & 4.6 & 4.2 & 5.0 \\
    $p_t$    (GPa)     & 8.3 &  & 13.9  \\
    \end{tabular}
  \end{ruledtabular}
\end{table}

\begin{figure}
  \includegraphics[width=0.5\textwidth]{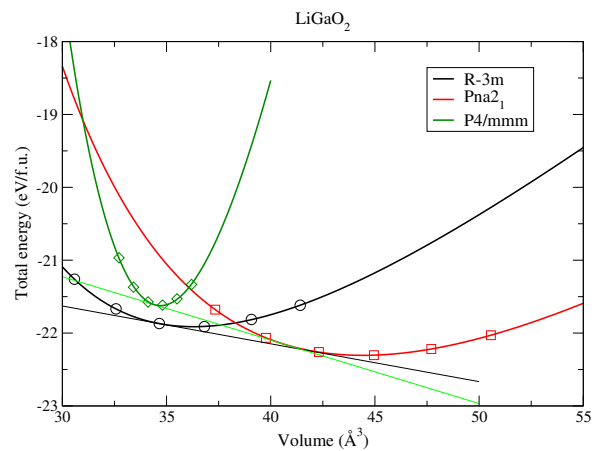}
  \caption{Energy volume curves of LiGaO$_2$ in $R\bar{3}m$, $P4/mmm$
    and $Pna2_1$ phases and
    the common tangent constructions.  The data points are
    directly calculated, the lines are Murnaghan equation of state fits.
    \label{figevollgo}}
\end{figure}

We calculated the total energies of the
$Pna2_1$, $R\bar{3}m$ and $P4/mmm$ phases as function of volume, keeping
the ratios of the lattice constants (shape of the cell) and internal parameters fixed
within the GGA-PBE functional with the FP-LMTO method. In order to compare
the energies accurately, we used the same muffin-tin radii for the three
structures,
chosen to be touching in the $Pna2_1$ structure at 0.94 compression of the lattice constants.
This avoids overlap of the spheres in the other structures. We included augmented plane waves
in the basis set up to 3 Ry and used a large basis set $spdf-spd$ on Li and Ga and also
included the Ga-$3d$ semi-core states as bands. We found the energy difference between the $R\bar{3}m$ and $Pna2_1$ 
phases at their equilibrium volume to be converged to 0.1 eV by increasing the cut-off of the
augmented plane waves to 4 Ry. 

The energy-volume curves and common tangent constructions are shown in
Fig.\ref{figevollgo}. The energy volume curves were fitted
to the Murnaghan equation of
state and the latter was used to determine the
transition pressure from the common-tangent rule or equivalently,
setting the enthalpy, $H_i(p)=E_i[V_i(p)]+pV_i(p)$
equal for two different phases $i$. 
The  equation of state fitted parameters are given in Table \ref{tabeos}. The energy difference
between the $Pna2_1$ and $R\bar{3}m$ phases
amounts to 97.5 meV/atom which is somewhat larger than the value
given in MP\cite{MP}  of 61 meV. The $P4/mmm$ phase is found to
have higher energy than the $R\bar{3}m$, namely 170 meV/atom higher
than the $Pna2_1$ phase but occurs at minimum energy volume of 34.7 \AA$^3$
close to that of the $R\bar{3}m$ phase. 
The bulk moduli might be somewhat overestimated because we did not allow the
structure to relax at each volume. Prior work found a bulk modulus of $\sim$95 GPa for the $Pna2_1$
structure\cite{Boonchun10,Lei13} a 142.29 GPa in $R\bar{3}m$.\cite{Lei13}

Phase transitions of the $\beta$-LiGaO$_2$ $Pna2_1$ to the $\alpha$
form $R\bar{3}m$ and other
disordered rocksalt type forms have been studied by Raman spectroscopy.\cite{Lei13,Lei10}
At 14 GPa a transition is found to the $I4/m$ structure, which is closely related to the rocksalt structure. This is very close to our value of the transition
to the $P4/mmm$ phase. 
According to Lei \etal \cite{Lei13}, the $\alpha$-phase can be prepapred from the $\beta$-phase at 7 GPa
and 1000$^\circ$C. This is compatible with the transition pressure found here
between the $Pna2_1$ and $R\bar{3}m$ phase of 8.3 GPa. From our convergence
studies we estimate our value to have an uncertainty of a few GPa, in particular
because we calculated our energy-volume curves without relaxing the
internal parameters or shape of the cell. Our calculations predict that
the transition to $R\bar{3}m$ should occur first but kinetic reasons
could prevent this and once at higher pressure, the transition
may then proceed already to the rocksalt type phase.\cite{Lei13}

\subsection{Band structure of LiGaO$_2$ in $Pna2_1$ structure}
For our band structure investigations of
LiGaO$_2$ in the $Pna2_1$ structure, we used the experimental
lattice parameters and atomic positions from Marezio \etal\cite{Marezio65}.
For completeness they are given in Table \ref{tabstruc}.

\begin{figure}
  \includegraphics[width=8cm]{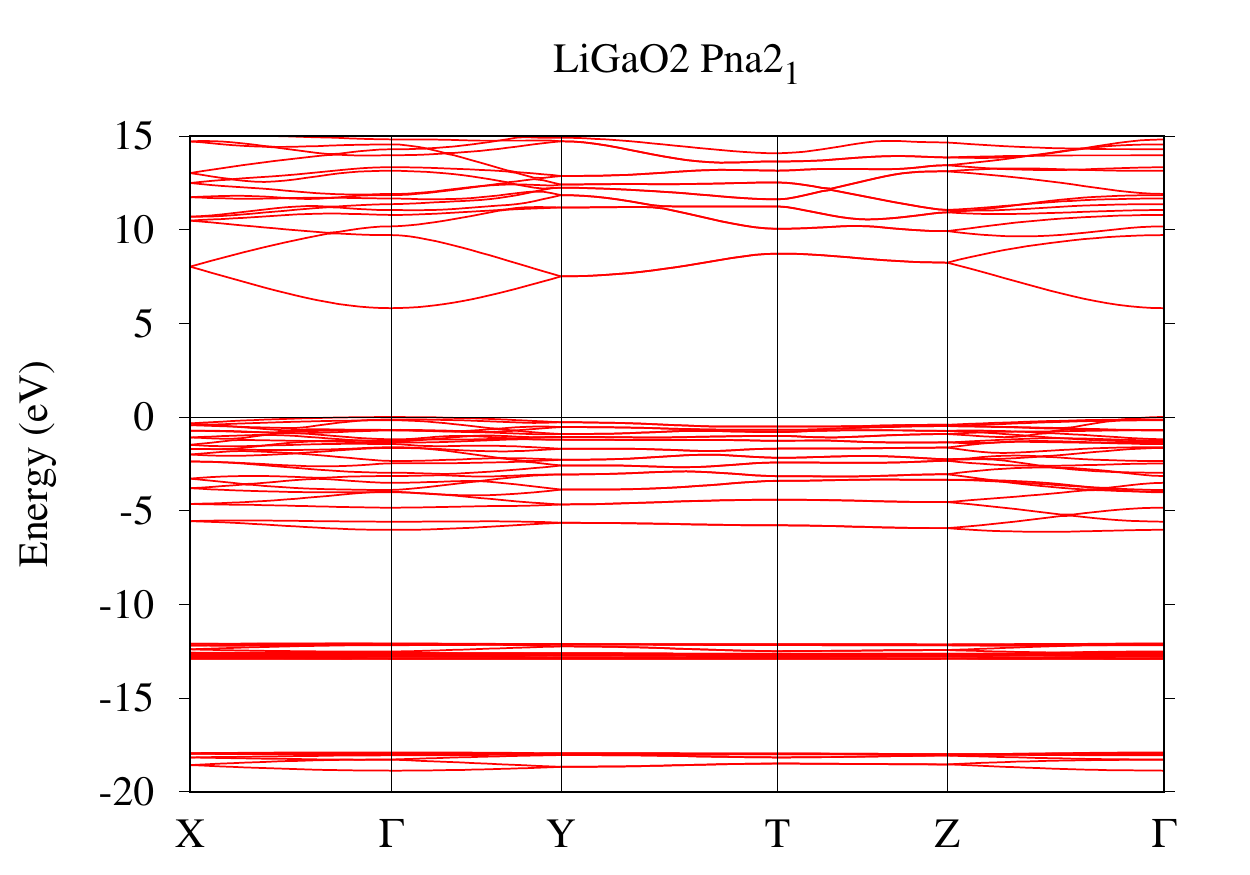}
  \caption{Band structure LiGaO$_2$ in
    $Pna2_1$ structure in 0.8$\Sigma$ QS$GW$ approximation.\label{figband}}
\end{figure}

\begin{figure}
    \includegraphics[width=0.5\textwidth]{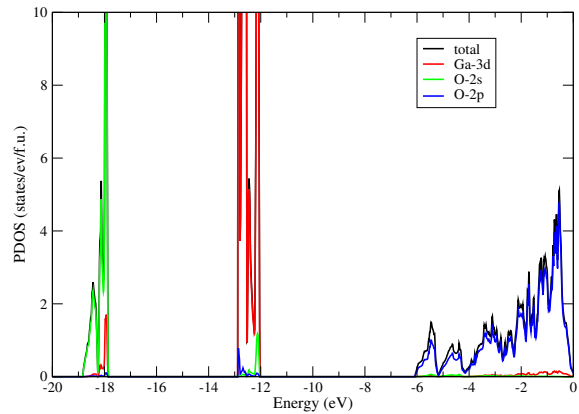}
    \includegraphics[width=0.5\textwidth]{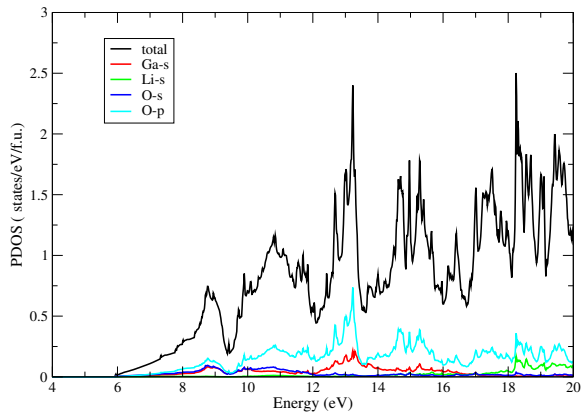}
       \caption{Total and partial densities of states for valence band (top)
     and conduction band (bottom) region.\label{figpdos}}
\end{figure}

\begin{figure}
  \includegraphics[width=0.5\textwidth]{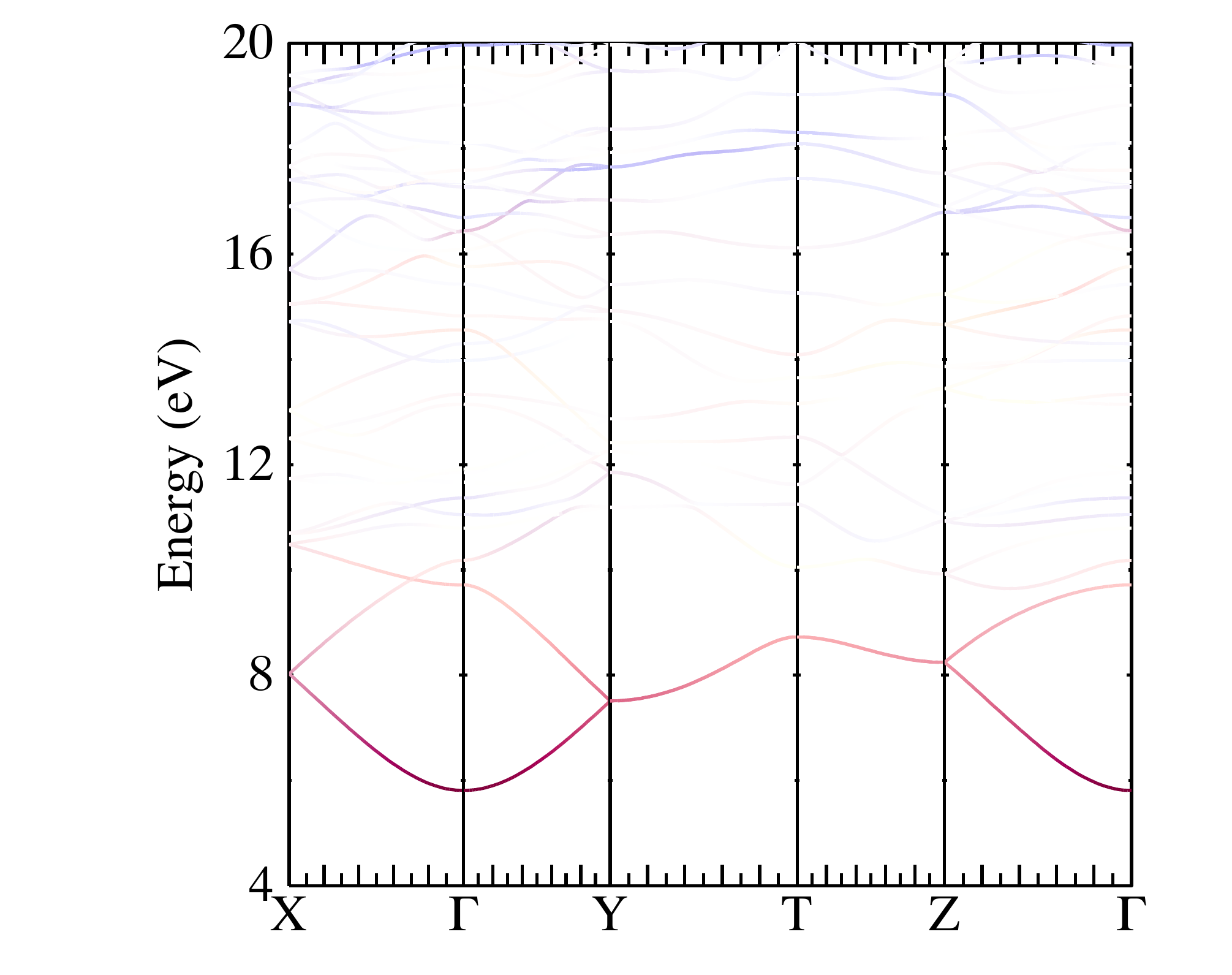}
  \caption{Conduction band structure of LiGaO$_2$ in $Pna2_1$
    structure in $0.8\Sigma$ approximation showing the Ga-$4s$ in red
    and Li-$2s$ in blue contributions to the bands.\label{figbndlgocol}}
\end{figure}

The band structure of LiGaO$_2$ in QSGW $0.8\Sigma$ is shown on a
  large energy scale in Fig. \ref{figband}. The density of states 
  resolved in various partial densities
  of state (PDOS) are shown in Fig. \ref{figpdos} both in the
  conduction band and the valence band region. This shows that
  the lowest set of narrow bands at about $-$19 eV  are the O-$2s$
  derived bands and the ones at about $-$12 eV below the VBM are the
  Ga-$3d$ derived bands.  The next set of bands between $-6$ eV and 0 eV
  are the O-$2p$ dominated band. In the conduction bands we see also significant
  oxygen contributions because these are antibonding bands.  It shows
  that the conduction band minimum has more Ga-$4s$ contribution and
  the main Li-$2s$ only occurs at significantly higher energy, above 18 eV.
  This is consistent with the high electropositivity of Li.  This
  can also be seen in the colored band plot in Fig. \ref{figbndlgocol}.

\begin{figure}
    \includegraphics[width=0.5\textwidth]{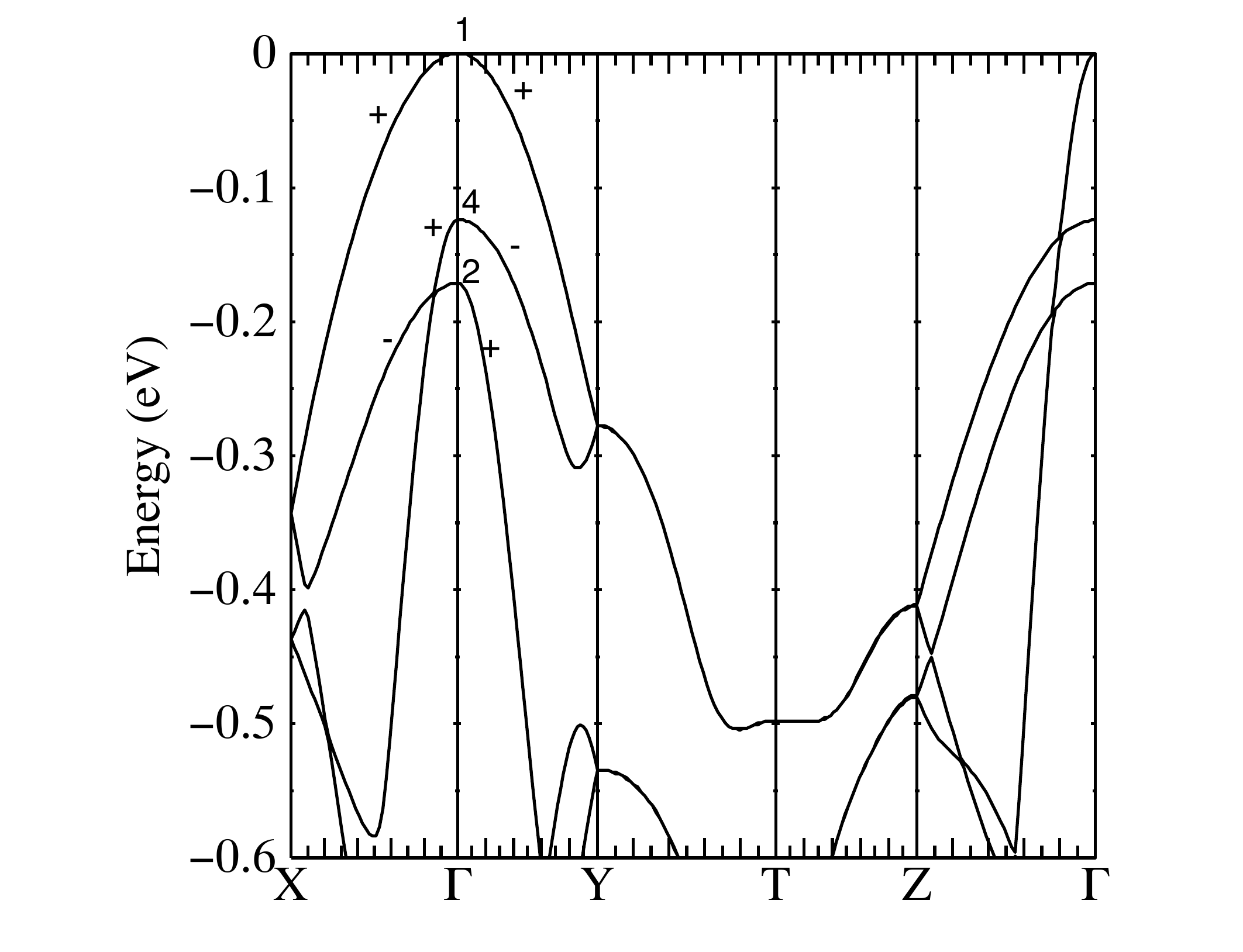}
    \includegraphics[width=0.5\textwidth]{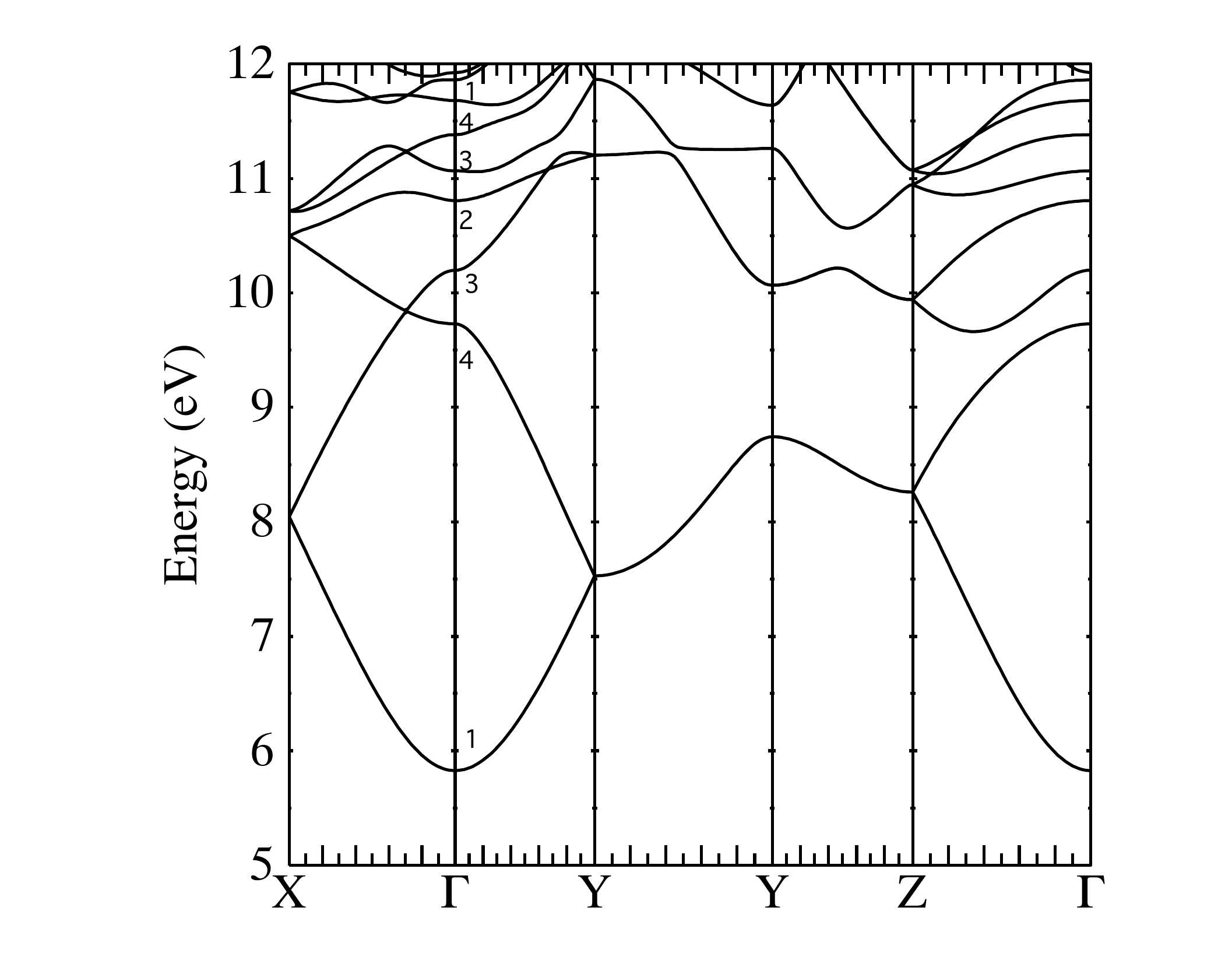}
    \caption{Symmetry labeled bands. Please note different
      scales in (top) near valence band maximum) and (bottom)
    conduction band region.\label{figbndzoom}}
\end{figure}

\begin{table}
  \caption{Character table of point group $C_{2v}$ indicating
    both the chemistry and Koster notation of the irreducible representations.
    \label{tabchar}}
  \begin{ruledtabular}
    \begin{tabular}{ll|rrrr|l}
      chem&Koster& $1$ & $2_{1z}$ & $n_x$ & $a_y$ & functions\\\hline
      $a_1$ & $\Gamma_1$ & 1 & 1 & 1 & 1 & $z$, $x^2$,$y^2$,$z^2$ \\
      $a_2$ & $\Gamma_3$ & 1 & 1 & $-$1 & $-$1 & $xy$ \\
      $b_1$ & $\Gamma_4$ & 1 & $-$1 & $-$1 & 1 & $x$, $xz$ \\
      $b_2$ & $\Gamma_2$ & 1 & $-$1 & 1 & $-$1 & $y$, $yz$ \\
    \end{tabular}
  \end{ruledtabular}
\end{table}

Next we show a zoom in on the region near the conduction band minimum
and valence band maximum in Fig. \ref{figbndzoom}.
The bands are symmetry labeled according to the character table
given in Table \ref{tabchar}, following the Koster \etal \cite{Koster} notation.
The band gap is direct at $\Gamma$ and is 5.81 eV.
The conduction band minimum (CBM) is found at $\Gamma$ and shows a strongly
dispersive band with low effective mass as is typical for cation-$s$-like
CBMs.  The valence band maximum (VBM) is also at $\Gamma$ but shows
crystal field splitting compared to the three-fold degenerate $p$-like
state seen in cubic zincblende materials.  Compared to wurtzite
which has only $z$ split from $x,y$ states, there is here a full splitting
in three levels even without  spin-orbit coupling.

We can see that the CBM has $\Gamma_1$ symmetry, consistent with its dominant
$s$-like character.  The VBM also has $\Gamma_1$ symmetry, separated by
124 meV from the $\Gamma_4$ band and the latter separated by 48 meV from
the next $\Gamma_2$.  Since $\Gamma_1,\Gamma_4,\Gamma_2$ correspond to $z,x,y$
respectively, this implies that optical transitions from the $\Gamma_1$ VBM
to the CBM are dipole allowed for ${\bf E}\parallel {\bf c}$, from the
$\Gamma_4$ band for ${\bf E}\parallel {\bf a}$ and for the $\Gamma_2$
band for ${\bf E}\parallel {\bf b}$. These lead to different onsets of
absorption as confirmed by our calculations of the optical response
discussed below.

Along the $\Gamma-X$ direction, only the $a_y$ mirror symmetry leaves the
${\bf k}$-point invariant and the states can thus be labeled as even, $+$, or  odd,
$-$, with respect to that mirror plane. From the character table, one
can easily see that $\Gamma_1$ and $\Gamma_4$ are compatible with $+$
and $\Gamma_2$, $\Gamma_3$ are compatible with $-$ along this direction.
Therefore the second and third valence band (counting down from the top)
are allowed to cross along $\Gamma-X$. Likewise along $\Gamma-Y$, only
the $n_x$ mirror survives. The compatibility relations are now
$\{\Gamma_1,\Gamma_2\}\rightarrow+$, $\{\Gamma_3,\Gamma_4\}\rightarrow-$.
On the other hand, along $\Gamma-Z$ the group of ${\bf k}$ stays $C_{2v}$
and hence the symmetry labeling at $\Gamma$ also applies along the $\Gamma-Z$ axis. Since the top three bands have different symmetry label they are allowed
cross along this line.

In the conduction band  we have only labeled the bands at $\Gamma$ since
the labels along the three orthogonal directions from $\Gamma$ is already
clear from the previous paragraph. 
We can see that there is a large gap of 3.90 between the CBM and next
conduction band (CBM2), which is of $\Gamma_4$ symmetry. This is favorable
for transparent conductor applications since only light with photon energy
larger than 3.9 eV (wavelength $\lambda<318$ nm) and with polarization
${\bf E}\parallel{\bf a}$ would be absorbed by electrons
near the conduction band minimum introduced by n-type doping.
The next CBM3 has $\Gamma_3$ symmetry which is dipole forbidden
for any light polarization. The first allowed transitions from the
CBM for ${\bf E}\parallel{\bf b}$ would only occur to the CBM4 band of $\Gamma_2$ symmetry at 4.98 eV. The lowest  conduction band at $\Gamma$ of $\Gamma_1$
symmetry is CBM7 at 5.85 eV.

\begin{table}
  \caption{Effective masses (in units of the free electron mass)
    in $Pna2_1$ LiGaO$_2$ and energy levels at $\Gamma$ relative to the VBM. \label{tabmass}}
  \begin{ruledtabular}
    \begin{tabular}{cccccc}
      band& irrep& $E$ (eV) & $m_x$ & $m_y$ & $m_z$ \\ \hline
      CBM  &  $\Gamma_1$ & 5.81     & 0.39 & 0.39 & 0.41 \\
      VBM1 & $\Gamma_1$  & 0        & 3.85 & 3.50 & 0.42 \\
      VBM2 & $\Gamma_4$  & $-0.124$ & 0.45 & 3.50 & 3.80 \\
      VBM3 & $\Gamma_2$  & $-0.172$ & 3.15 & 0.58 & 3.80 \\ \hline
          \end{tabular}
  \end{ruledtabular}
\end{table}

Because of the orthorhombic symmetry the mass tensors
at each band at $\Gamma$ are diagonal with a different
mass in each of the $x,y,z$ directions.
These are given in Table \ref{tabmass}.
One can see that the conduction band mass is small and close to isotropic.
The valence bands each have one light mass and and two heavy mass directions.
The light mass is in the direction corresponding to the symmetry
of the band, for example it is in the $x$ direction for the $\Gamma_4$
band, in the $y$ direction for the $\Gamma_2$ band and in the $z$ direction
for the $\Gamma_1$ band. 

%\textcolor{red}{Following Ref. \onlinecite{Punya11} these can be used to
%derive the parameters of a generalization of the Kohn-Luttinger
%Hamiltonian for these orthorhombic crystals.  While some of
%the parameters in this Hamiltonian can be obtained within a quasicubic
%or quasihexagonal approximation, a more reliable
%value of these parameters is obtianed by including  a parabolic
%band fit along directions intermediate between the orthorhombic
%crystal directions in the overall fit.}

We also performed calculations including spin-orbit coupling.
The valence band maximum is then still split in three levels and
the difference from the calculation without spin-orbit coupling was found
to be negligible. In  the parent compound ZnO, spin-orbit splitting
plays an  important role because of the antibonding contribution from the
Zn-$3d$ orbitals in the VBM. In fact, it leads to an effectively  negative
spin-orbit splitting parameter of the VBM in that case.\cite{Lambrecht02}
In LiGaO$_2$,
however the Ga-$3d$ orbitals lie significantly lower, reducing this negative
contribution and hence apparently almost completely cancelling the positive
but already small contribution of the O-$2p$ orbitals.  
%Other off-diagonal parameters in this
%Hamiltonian are related to the avoided crossings of bands of the
%same symmetry along the various symmetry directions.
%\textcolor{red}{A full set of effective Hamiltonian parameters are provided in
%Table\ref{tabKLparam}. to be done}

\begin{table}
  \caption{Dielectric constants $\varepsilon^\infty$ of LiGaO$_2$ in
    various approximations.\label{tabdielconst}}
  \begin{ruledtabular}
    \begin{tabular}{ldddd}
       method & \multicolumn{1}{c}{$\varepsilon^\infty_x$} &  \multicolumn{1}{c}{$\varepsilon^\infty_x$} &  \multicolumn{1}{c}{$\varepsilon^\infty_x$} & \multicolumn{1}{c}{$(\varepsilon^\infty_x\varepsilon^\infty_y\varepsilon^\infty_z)^{1/3}$} \\ \hline
      LDA\cite{Boonchun10} Berry       & 3.492 & 3.342 & 3.490 & 3.441 \\
      GGA                              & 2.803 & 2.750 & 2.824 & 2.792 \\
      0.8$\Sigma$ $d\Sigma/dk$         & 4.987 & 5.421 & 7.256 & 5.810 \\
      0.8$\Sigma$ $d\Sigma/dk=0$       & 1.818 & 1.796 & 1.828 & 1.810 \\
      0.8$\Sigma$ rescaling            & 2.445 & 2.399 & 2.462 & 2.435 \\
      Expt.  \cite{Tumenasi17}         & 2.99  & 2.90  & 2.99  & 2.960 \\
      Expt.  \cite{Nanamatsu72}        & 3.05  & 2.99  & 3.05  & 3.030 \\
    \end{tabular}
  \end{ruledtabular}
\end{table}

The optical response functions were calculated within the long-wavelength
limit and independent particle approximation. In other words, they
include vertical band-to-band transitions including the dipole
matrix elements but no local field  or excitonic effects. In this
case the imaginary part of the dielectric function
$\varepsilon_2(\omega)$ is given by 
\begin{eqnarray}
\varepsilon_2(\omega)&=&\frac{8\pi^2e^2}{V \omega^2}\sum_n\sum_{n'}\sum_{{\bf k}\in BZ}f_{n{\bf k}}(1-f_{n'{\bf k}}) \nonumber \\
&&|\langle \psi_{n{\bf k}}|[H,{\bf r}]|\psi_{n'{\bf k}}\rangle|^2\delta(\omega-\epsilon_{n'{\bf k}}+\epsilon_{n{\bf k}}), 
\end{eqnarray}
with $\epsilon_{n{\bf k}}$ the band eigenvalues, $\psi_{n{\bf k}}$
the Bloch eigenstates, $f_{n{\bf k}}$ the Fermi function occupation factors
of these states. It  uses matrix elements of the velocity operator
$\dot{\bf r}=(i/\hbar)[H,{\bf r}]$, which differ from the momentum
matrix elements ${\bf p}/m$ because of the non-local contribution of the
self-energy operator to the Hamiltonian. 
The real part $\varepsilon_1(\omega)$ is obtained from a Kramers-Kronig
transformation and the optical absorption coefficient is given by 
$\alpha(\omega)=2\varepsilon_2(\omega)/n(\omega)$
with  $n(\omega)$ the index of refraction given by
$\tilde n(\omega)=\sqrt{\varepsilon_1(\omega)+i\varepsilon_2(\omega)}=n(\omega)+i\kappa(\omega)$. 

\begin{figure}
  \includegraphics[width=9cm]{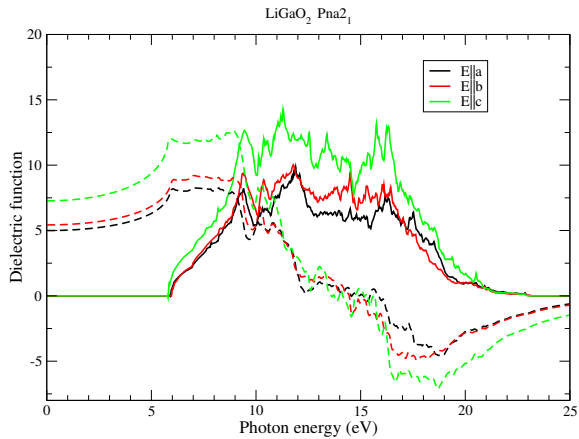}
  \caption{Imaginary $\varepsilon_2$  and real $\varepsilon_1$ parts
    of the optical dielectric function for $x,y,z$ directions.\label{figeps}}
\end{figure}

\begin{figure}
  \includegraphics[width=8cm]{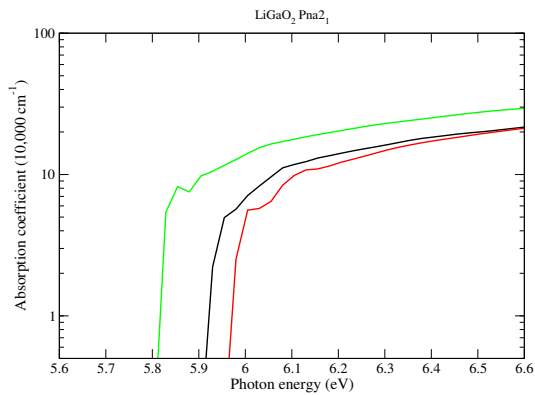}
  \caption{Optical absorption onset $\alpha(\omega)$ for the three
    crystal directions.\label{figabs}}
\end{figure}

The imaginary and real parts of the dielectric function are shown in
Fig. \ref{figeps} and the absorption coefficient is  shown on 
a log scale near the onset of absorption in Fig. \ref{figabs}.
It shows the different absorption onsets for different polarization
consistent with the symmetry analysis present above.

The values of $\varepsilon_1(\omega=0)_{\alpha\alpha}$  correspond
to the $\varepsilon^\infty$ and can be compared with the results obtained
from a Berry-phase calculation in LDA in Boonchun and Lambrecht.\cite{Boonchun10} This comparison is given in Table \ref{tabdielconst} along
with our values both based on the GGA and the QS$GW$ 0.8$\Sigma$ band
structure. The Berry phase calculations in principle includes local
field corrections which tend to reduce the value by about 5 \%.
In spite of this our GGA values are smaller than the
ones in Ref. \onlinecite{Boonchun10}.  The increase in gap due to
QS$GW$ is expected to reduce the dielectric constant
(because energy denoninators in the expression for $\varepsilon$
are increased) when applied in a naive way 
without taking into account the matrix element rescaling,
from the non-local contribution to the velocity operator from the self-energy.
This is indicated in the table by the $d\Sigma/dk=0$. However when
including the $d\Sigma/dk$ contribution the dielectric constant
is in fact increased.  Instead of  using the explicitly calculated
$d\Sigma/dk$ one can also use the rescaling rule proposed by
Levine and Allan \cite{LevineAllan89} in the context of a scissor correction.
One then rescales the matrix elements by a factor
$(\epsilon_{n{\bf k}}-\epsilon_{{n^\prime\bf k}})/(\epsilon^{GGA}_{n{\bf k}}-\epsilon^{GGA}_{n^\prime{\bf k}})$. This apparently restores values close to the
GGA.
Finally, we compare to experimental values. Interestingly, the GGA results
seem to be in better agreement with the experiment but we should caution
that this could be due to a compensation of errors because we did
not include local field or excitonic effects here.

\begin{table}
  \caption{Band gap of LiGaO$_2$ in $Pna2_1$ structure in different
    approximations.\label{tabgap}}
  \begin{ruledtabular}
    \begin{tabular}{lcccccc}
                        & GGA & QS$GW$ & 0.8$\Sigma$ & $+ZPM$ & Expt. \\ \hline
      At expt. $a,b,c$  & 3.363 & 6.363   & 5.81 & 5.6 & 5.26-5.5 \\
      at PBE $a,b,c$    & 3.201 & 6.245 & 5.69 & 5.5 & \\
    \end{tabular}
  \end{ruledtabular}
\end{table}

Let us now focus on the band gap in Table\ref{tabgap}.
First, in terms of {\bf k}-convergence of the $\tilde\Sigma({\bf k})$,
the $0.8\Sigma$ gap obtained with a $3\times3\times3$ mesh is 5.804 eV while that with a $4\times4\times4$ mesh it is 5.812 eV, showing that we
have reached convergence to 0.01 eV in terms of {\bf k}-convergence.
The full QS$GW$ gaps are 6.35 and 6.36 eV  with $3\times3\times3$ and $4\times4\times4$ respectively. 
One can see that the $GW$ corrections to the GGA band structure are
significant. Also, we find a gap of 5.8 eV in fair agreement with experiments
when adding a 0.8 correction factor to the $\tilde\Sigma-v_{xc}^{DFT}$.
This is known as the 0.8$\Sigma$ approximation. It takes into
account that within QS$GW$, the polarization propagator
does not include electron-hole interaction effects, or ladder
diagrams, which typically
leads to an underestimate of the screening. Although including
such effects has recently become possible in the Bethe-Salpeter-Equation
(BSE) approach\cite{Cunningham18}, it is still prohibitively time and
memory consuming for
as system like the present one with 16 atoms per cell.  On the other
hand for many known cases, this under screening by the random
phase approximation (RPA) was found to be about 20 \% as
documented in Ref. \onlinecite{Churna18}. Hence the commonly used
0.8$\Sigma$ correction factor.

While a full calculation of the zero point motion correction to the gap
by electron phonon coupling is time consuming, we can make at least
some estimate of this effect following the approach of
Ref. \onlinecite{Lambrecht17}. In a highly ionic material, the main effect
comes from the lattice polarization correction (LPC), which is
essentially the polaronic shift of the band edges. Its origin
can be viewed as the contribution to the screening from
the lattice polarization in the long-wavelength limit. 
For a material with a single LO phonon, this shift
is given by\cite{Lambrecht17}
\begin{equation}
  \Delta E_P=-\alpha_P \hbar\omega_{LO}/2=\frac{e^2}{4a_P}\left[ \frac{1}{\varepsilon_\infty}-\frac{1}{\varepsilon_0}\right]
\end{equation}
where $a_P=\sqrt{\hbar/(2\omega_{LO}m_*)}$ is a polaron length
defined in terms of the effective mass of the electrons for the conduction
band shift and holes for the valence band shift and $\omega_{LO}$ is the longitudinal optical phonon. The dimensionless polaronic coupling factor is
$\alpha_P$ and the dielectric constants at frequencies high above
the phonons but well below the gap is $\varepsilon_\infty$ and
the static dielectric constant below the phonon modes is $\varepsilon_0$.
The factor 1/2 in this equation was obtained by applying a cut-off
to the wave vector  of order $1/a_P$ while the classic Fr\"ohlich
estimate of the polaronic shift  does not include such
a cut-off.
We obtain an upper limit to this correction by calculating the contribution
from the highest energy optical phonon.  The phonons in LiGaO$_2$
were calculated in Boonchun \etal\cite{Boonchun10} and this paper
also provides values for the dielectric constants needed here.
Averaging over the $b_{1L}$, $b_{2L}$ and $a_{1L}$ phonons, we take
an estimate of $\omega_{LO}\approx 750$ cm$^{-1}$. Averaging the dielectric
constants over directions, we obtain $\varepsilon_\infty\approx3.44$,
$\varepsilon_0\approx6.93$. The factor $[\varepsilon_\infty^{-1}-\varepsilon_0^{-1}]$ then amounts to 0.146. Now using a hole mass of about $m_*=3$
we obtain $a_P^h\approx 7 a_0$ with $a_0$ the Bohr radius.  The expected
shift is then about 0.14 eV. The conduction band shift should be significantly
smaller because the electron effective mass is only 0.33 giving an $a_P^e\approx 21 a_0$, giving a shift of 0.05 eV at most. This gives an estimated
gap correction of $-$0.2 eV form the highest frequency LO phonons.
When multiple infrared active phonons are present, each phonon has
a separate contribution to the $\varepsilon_0/\varepsilon_\infty$ factor
according to the Lyddane-Sachs-Teller relation but also each long-range
Fr\"ohlich type electron-phonon
coupling parameter depends on the eigenvector of the phonon.\cite{Verdi15}
It does becomes much more difficult to make an estimate. The above should
be viewed only as an order of magnitude estimate of the effect.
Including this estimate of a negative shift of a few 0.1 eV
of the electron  phonon coupling zero-point motion
correction, our gap is in excellent agreement with the experimental value.
The remaining discrepancy of our value with the most recent
experimental determination of 5.26 eV is probably
due to this measurement being done at room temperature. 

It is also of interest to check how the QS$GW$ correction of the gap
is split over the valence and conduction band edges. We find in the $0.8\Sigma$
approximation that the VBM shifts down from the GGA value by 1.321 eV
while the CBM shifts up by 1.128 eV, thus giving a gap correction of 2.449
eV. This is related to the Ga-$3d$ contribution to the VBM. The Ga-$3d$
bands are found to shift down by about 2 eV in the $0.8\Sigma$ approximation
compared to GGA. This reduces the antibonding contribution of the Ga-$3d$
to the VBM and hence shifts down the VBM significantly.  The O-$2s$
bands also shift down by about 1.7 eV. 

We note that at the PBE calculated lattice constants, the
GGA gap is $\sim$0.12 eV smaller. This is because the larger lattice constant
typically leads to lower covalent interactions and hence
a lower band gap. Since volume in GGA
is about 3 \% overestimated, this allows us to estimate the
band gap deformation potential $dE_g/d\ln{V}\approx4$ eV.

\subsection{LiGaO$_2$ in $R\bar{3}m$ and $P4/mmm$ structures.}
\begin{figure}
  \includegraphics[width=8cm]{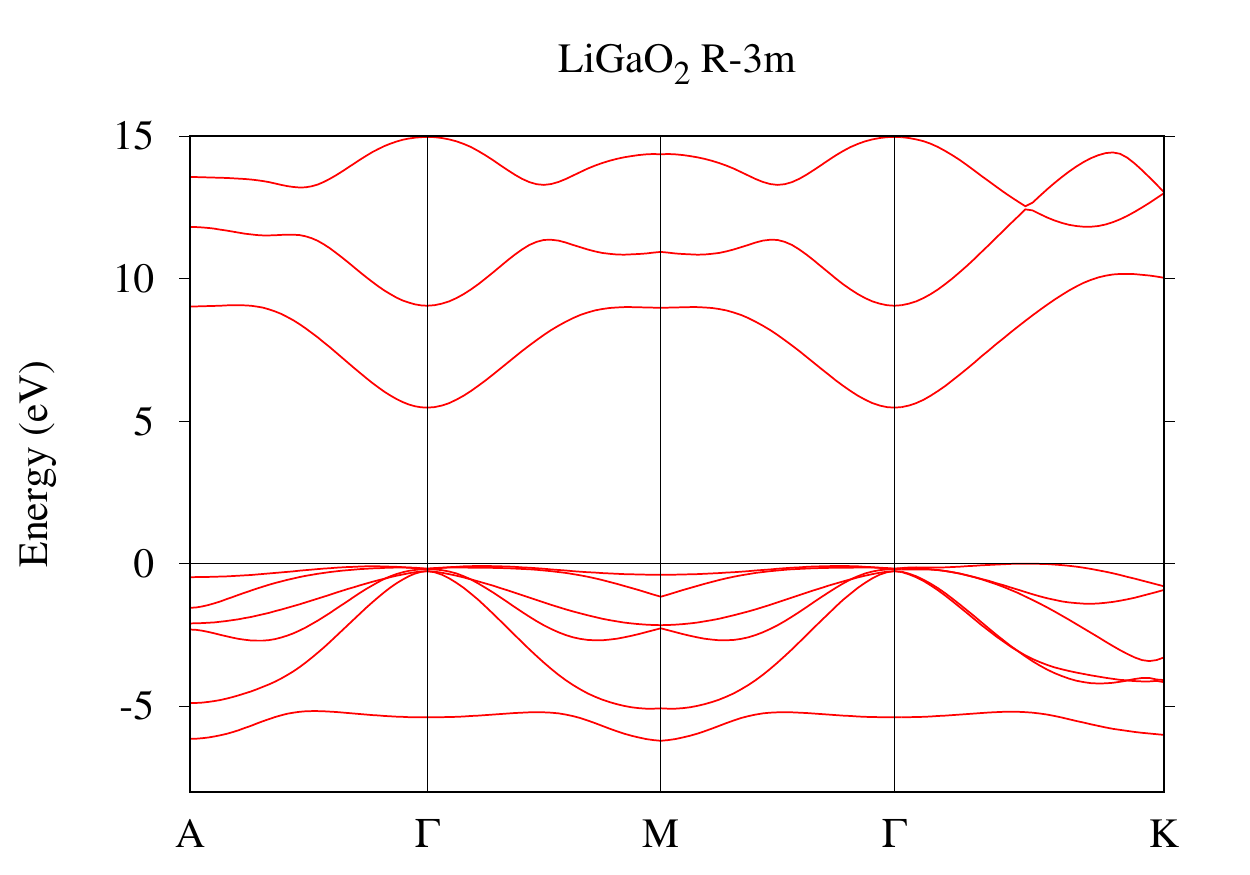}
  \caption{Band structure of LiGaO$_2$ in the $R\bar{3}m$ structure
    in $0.8\Sigma$ QS$GW$ approximation. \label{figlgor-3m}}
\end{figure}

The band structure of the $R\bar{3}m$ structure
of LiGaO$_2$ was calculated in the 0.8$\Sigma$ QS$GW$ approximation and is
shown in Fig. \ref{figlgor-3m}. Although we use here the primitive
cell of $R\bar{3}m$ we plot the bands along symmetry lines
of the conventional hexagonal cell, where $\Gamma-M-K$ lie in
the $k_z=0$ plane and $A$ lies at the Brillouin zone edge along the $k_z$ direction above $\Gamma$. This allows one to see the symmetries more easily.
We can see that the band gap
in this structure is slightly indirect, because the VBM occurs along
$\Gamma-K$ while the CBM remains at $\Gamma$. The direct gap at $\Gamma$ is
5.65 eV while the indirect gap is 5.47 eV.  The O-$2s$ and Ga-$3d$ bands
occur at about the same energies as in the $Pna2_1$ structure.

\begin{figure}
  \includegraphics[width=8cm]{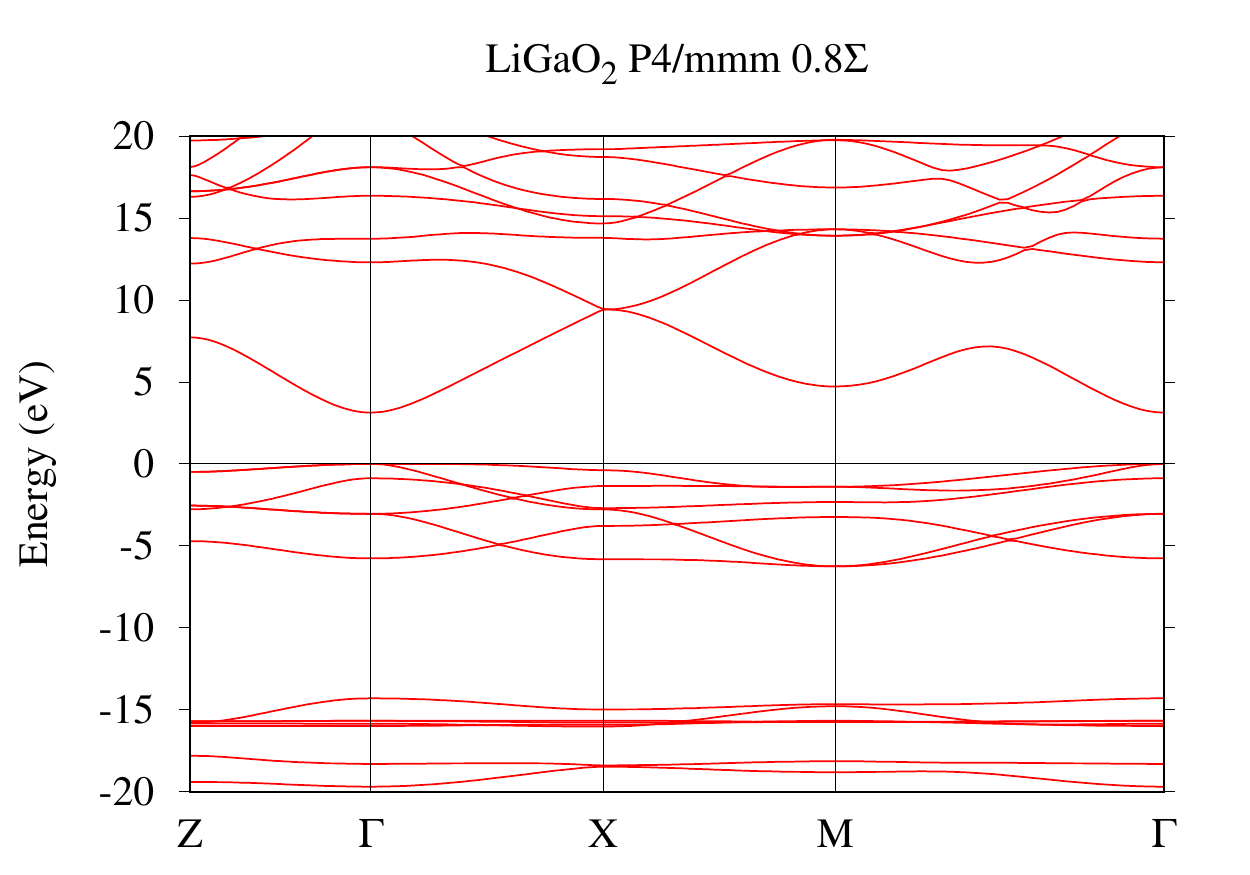}
  \caption{Band structure of LiGaO$_2$ in the $P4/mmmm$ structure in
    $0.8\Sigma$ QS$GW$ approximation.\label{figlgop4mmm}}
\end{figure}

The band structure in the $P4/mmm$ structure is shown in Fig \ref{figlgop4mmm}
in the $0.8\Sigma$ approximation. It is seen to have a direct gap but
with significantly smaller value of 3.129 eV.

\subsection{Structure and stability of NaGaO$_2$}
%\begin{table*}
%  \caption{Reduced coordinates, Wyckoff positions and symmetry operation linking equivalent sites, and lattice constant for NaGaO$_2$ in $Pna2_1$ structure optimized within GGA-PBE using abinit.\label{tabngostruc}}
%  \begin{ruledtabular}
%    \begin{tabular}{lcccc}
%      atom & Wyckoff & $x$ & $y$ & $z$ \\ \hline 
%      Na   & 4a      & 0.0725 & 0.6223 & $-$0.0112 ($-0.0131$)\\
%      Ga   & 4a      & 0.0642 & 0.1261 & 0.0019 (0.0000)\\
%      O$_\mathrm{Na}$ & 4a &0.1135 & 0.6615 & 0.0.4190 (0.4171) \\
%      O$_\mathrm{Ga}$ & 4a & 0.0421 & 0.0907 & 0.3474 (0.3455) \\ \hline
%      $a$ (\AA)  & $b$ (\AA) & $c$ (\AA) & $V$ (\AA$^3$) \\ \hline 
%      5.5716      & 7.1318 & 5.3288 & 211.743 \\ \hline
%      $2a/b$       &$2c/b$ & $b/2=a_w$& \\
%      1.5625      & 1.4944 & 3.5659 \\
%      $d_\mathrm{Na-O}^\parallel$ & $d_\mathrm{Ga-O}^\parallel$ \\
%      2.363 \AA & 1.875 \AA \\
%    \end{tabular}
%  \end{ruledtabular}
%\end{table*}

\begin{table*}
  \caption{Lattice constants, volume per formula unit, reduced coordinates,
    and bond lengths of NaGaO$_2$ in $Pna2_1$ structure optimized within GGA-PBE.\label{tabngostruc}}
  \begin{ruledtabular}
    \begin{tabular}{lcccc}
      $a$ (\AA)  & $b$ (\AA) & $c$ (\AA) & $V/fu$ (\AA$^3$) \\ \hline 
      5.6138      & 7.2377 & 5.3895 & 54.746 \\ \hline
      $2a/b$       &$2c/b$ & $b/2=a_w$& \\
      1.5513      & 1.4893 & 3.6189 \\ \hline
      atom & Wyckoff & $x$ & $y$ & $z$ \\ \hline 
      Na   & 4a      & 0.0717 & 0.6223 & 0.0128 \\
      Ga   & 4a      & 0.0630 & 0.1262 & $-$0.0005  \\
      O$_\mathrm{Na}$ & 4a &0.1149 & 0.6632 & 0.5822 \\
      O$_\mathrm{Ga}$ & 4a & 0.0400 & 0.08923 & 0.6559 \\ \hline
 &\multicolumn{4}{c}{bond lenths (\AA)} \\ \hline
       Ga-O$_\mathrm{Ga}^c$ & Ga-O$_\mathrm{Ga}^a$ & Ga-O$_\mathrm{Na}^a$ & Ga-O$_\mathrm{Na}^b$ \\ 
       1.875              & 1.864 & 1.880 & 1.875 \\
       Na-O$_\mathrm{Na}^c$ & Na-O$_\mathrm{Na}^a$ & Na-O$_\mathrm{Ga}^a$ & Na-O$_\mathrm{Ga}^b$ \\
       2.357               & 2.346               & 2.312             & 2.325 \
    \end{tabular}
  \end{ruledtabular}
\end{table*}

\begin{table}
  \caption{Structural parameters of NaGaO$_2$ in $R\bar{3}m$ structure. \label{tabstrucngor-3m}}
  \begin{ruledtabular}
    \begin{tabular}{cccc}
       & Na & Ga & O \\ \hline
      Wyckoff &  1a & 1b & 2c \\ 
      reduc. coord. & $(0,0,0)$ & $(\frac{1}{2},\frac{1}{2},\frac{1}{2})$ & $(\pm u,\pm u,\pm u)$ \\ \\ \hline
      $a=b=c$ (\AA) & $\alpha=\beta=\gamma$ & $V/fu$ (\AA$^3$) & $u$ \\ \hline
      5.675   & 30.892$^\circ$                & 42.742 & 0.2326 \\ \hline
      bond lengths (\AA) & Li-O & Ga-O \\ \hline 
                   & 2.389 & 2.047 \\
    \end{tabular}
  \end{ruledtabular}
\end{table}

\begin{table}
  \caption{Lattice constants and atomic positions of NaGaO$_2$
    in $P4/mmm$ structure.\label{tabstrucngop4mmm}}
  \begin{ruledtabular}
    \begin{tabular}{lcccc}
      atom & Wyckoff  & $x$ & $y$ & $z$  \\ \hline
      Na  & 1c       & 0.5    & 0.5     & 0.0  \\
      Ga  & 1b       & 0.0    & 0.5     & 0.0  \\
      O   & 1a       & 0.0    & 0.0     & 0.0 \\
      O   & 1d       & 0.5    & 0.5     & 0.5 \\ \hline
  lattice constants (\AA)    & $a=b$        & $c$ & $V/fu$  \\ \hline
     &  3.189       & 4.177  & 42.498 \\
  bond lengths (\AA)   & $\parallel c$ & $\perp c$ \\
   & 2.088 & 2.255    \\
    \end{tabular}
\end{ruledtabular}   
\end{table}

\begin{table}
  \caption{Cohesive energy ($E_0/f.u.$), equilibrium volume ($V_0$), bulk modulus ($B_0$)
    and its pressure derivative ($B_0^\prime$) and transition pressure, $p_t$ of
    NaGaO$_2$\label{tabeosngo}}
  \begin{ruledtabular}
    \begin{tabular}{lccc}
    property  & $R\bar{3}m$ &  $Pna2_1$ & $P4/mmm$\\ \hline
    $V_0$ (\AA$^3$)    & 41.19 & 53.35 & 42.50\\
    $E_0$ (eV/f.u.)    & 19.96 & 20.62 & 19.08\\
    $B_0$ (GPa)        & 161 & 116 & 1047 \\
    $B_0^\prime$        & 4.5 & 4.3 & 5.0 \\
    $p_t$    (GPa)     & 13.0 & & 40.0 \\
    \end{tabular}
  \end{ruledtabular}
\end{table}

\begin{figure}
  \includegraphics[width=0.5\textwidth]{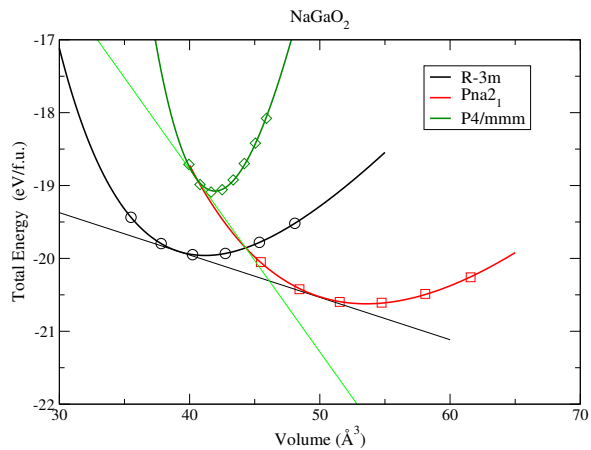}
  \caption{Total energy of NaGaO$_2$ (minus the energy of the corresponding
    free atoms) as function of volume for $Pna2_1$, $R\bar{3}m$ and $P4/mmm$
    structures
    and the common tangent constructions. Symbols are calculated, lines
    are Murnaghan equation of state fits. \label{figevngo}}
\end{figure}

The structural parameters of $Pna2_1$ NaGaO$_2$ were initially taken from
Materials Project and subsequently  relaxed within the GGA-PBE
approximation using the questaal and Quantum Espresso codes.
They are given in Table \ref{tabngostruc}.

We see that the $2a/b$ is significantly smaller than
the ideal value of $\sqrt{3}$, indicating that the $\gamma$ angle
between $a$ and $b$ has increased  from 120$^\circ$  to
129$^\circ$.
The Na-O bond lengths are  larger than the Li-O bond lengths while the
Ga-O bond length is about the same as in LiGaO$_2$.  As expected, the
volume per formula unit is somewhat larger than in LiGaO$_2$.

The optimized structural parameters of the $R\bar{3}m$ structure
are given in Table \ref{tabstrucngor-3m}.  We find this structure to
have a volume per formula unit that is 22\% smaller than the $Pna2_1$
structure and to be 165 meV/atom  higher in energy.
It is is thus again a high-pressure phase. Using the energy-volume
curves shown in Fig. \ref{figevngo}  we find a transition pressure of
13 GPa. This is comparable but slightly higher than in LiGaO$_2$.
Both are close to the 
transition from wurtzite ZnO to rocksalt ZnO,  which occurs at about 9 GPa. 
The $P4/mmm$ structure was optimized first as function of $c/a$
and then the equation of state was determined keeping the $c/a$ fixed.
The Murnaghan fit parameters and transition pressure from the $Pna2_1$
phase are given in Table \ref{tabeosngo}.

\subsection{Band structure of NaGaO$_2$ in $Pna2_1$ structure}
\begin{figure}
  \includegraphics[width=8cm]{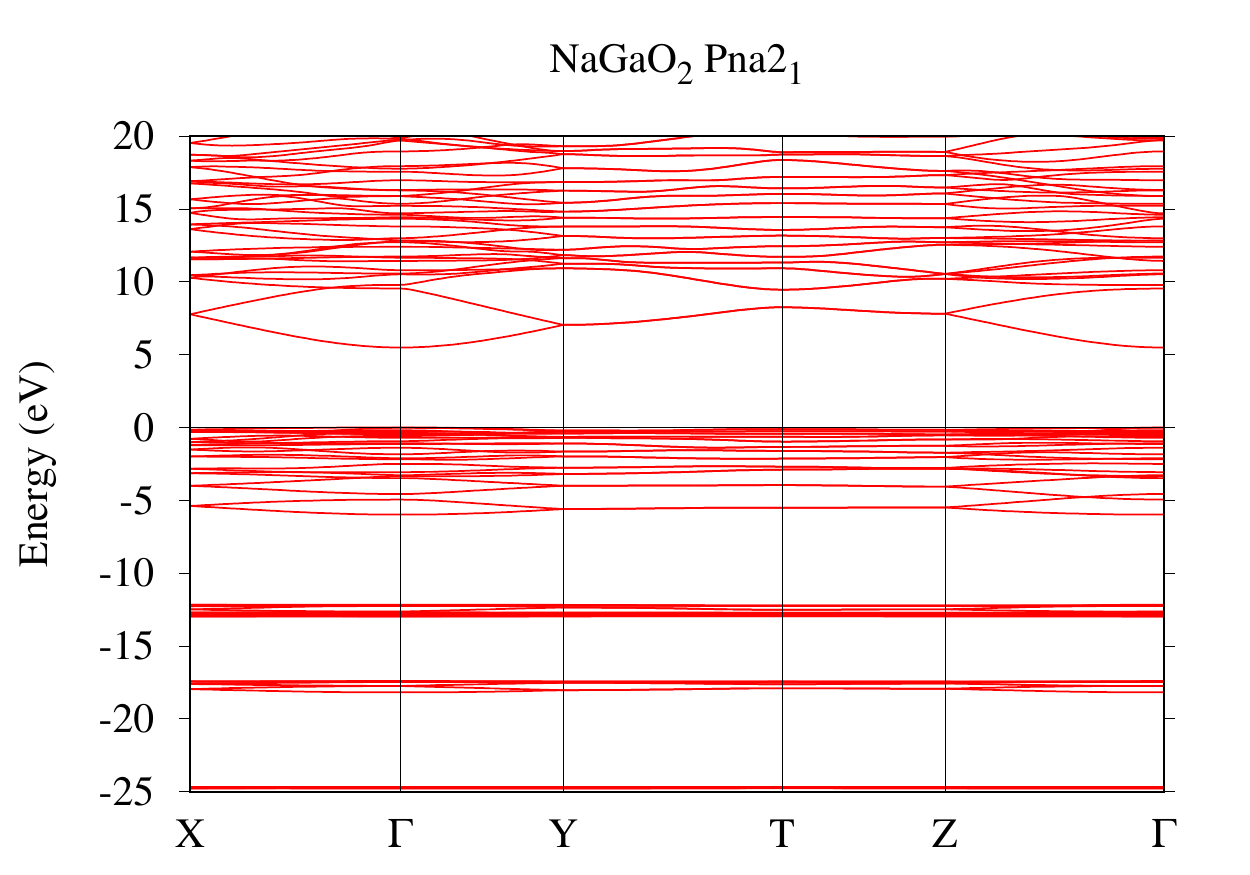}
  \caption{Band structure NaGaO$_2$ in
    $Pna2_1$ structure in 0.8$\Sigma$ QS$GW$ approximation.\label{figbandngo}}
\end{figure}

\begin{figure}
    \includegraphics[width=0.5\textwidth]{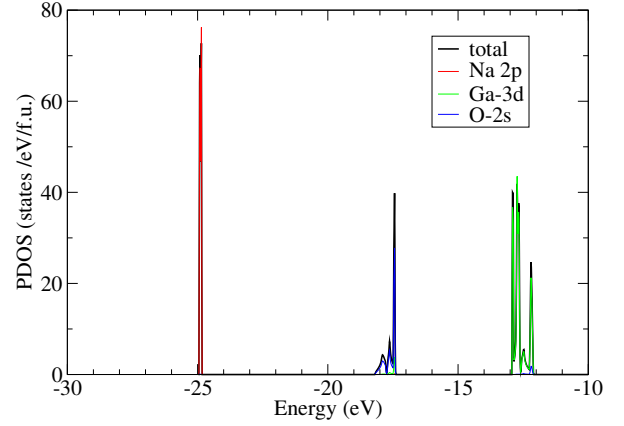}
    \includegraphics[width=0.5\textwidth]{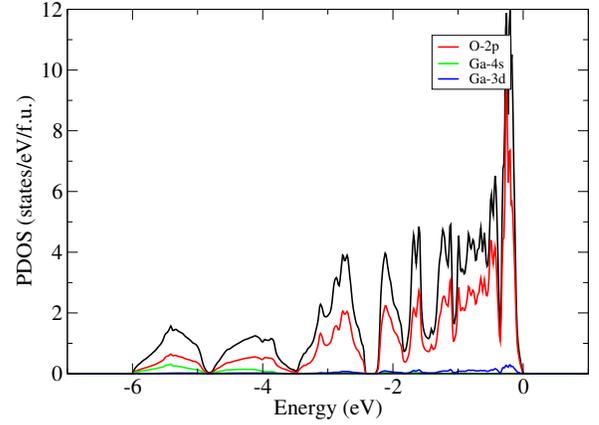}
    \includegraphics[width=0.5\textwidth]{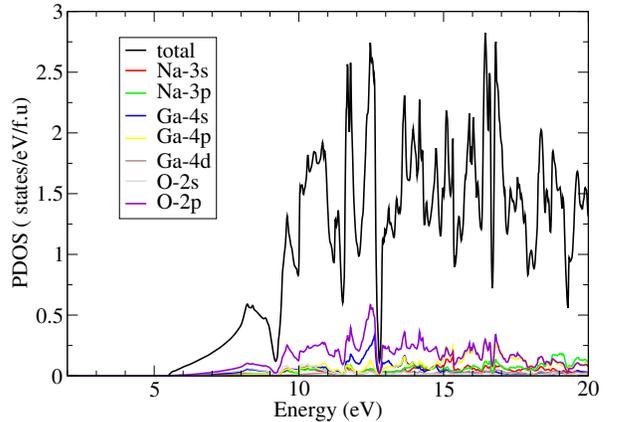}
    \caption{Total and partial densities of states for (top to bottom)
      semicore, valence band 
     and conduction band region in NaGaO$_2$ in $Pna2_1$ structure.\label{figpdosngo}}
\end{figure}
\begin{figure}[t]
  \includegraphics[width=9cm]{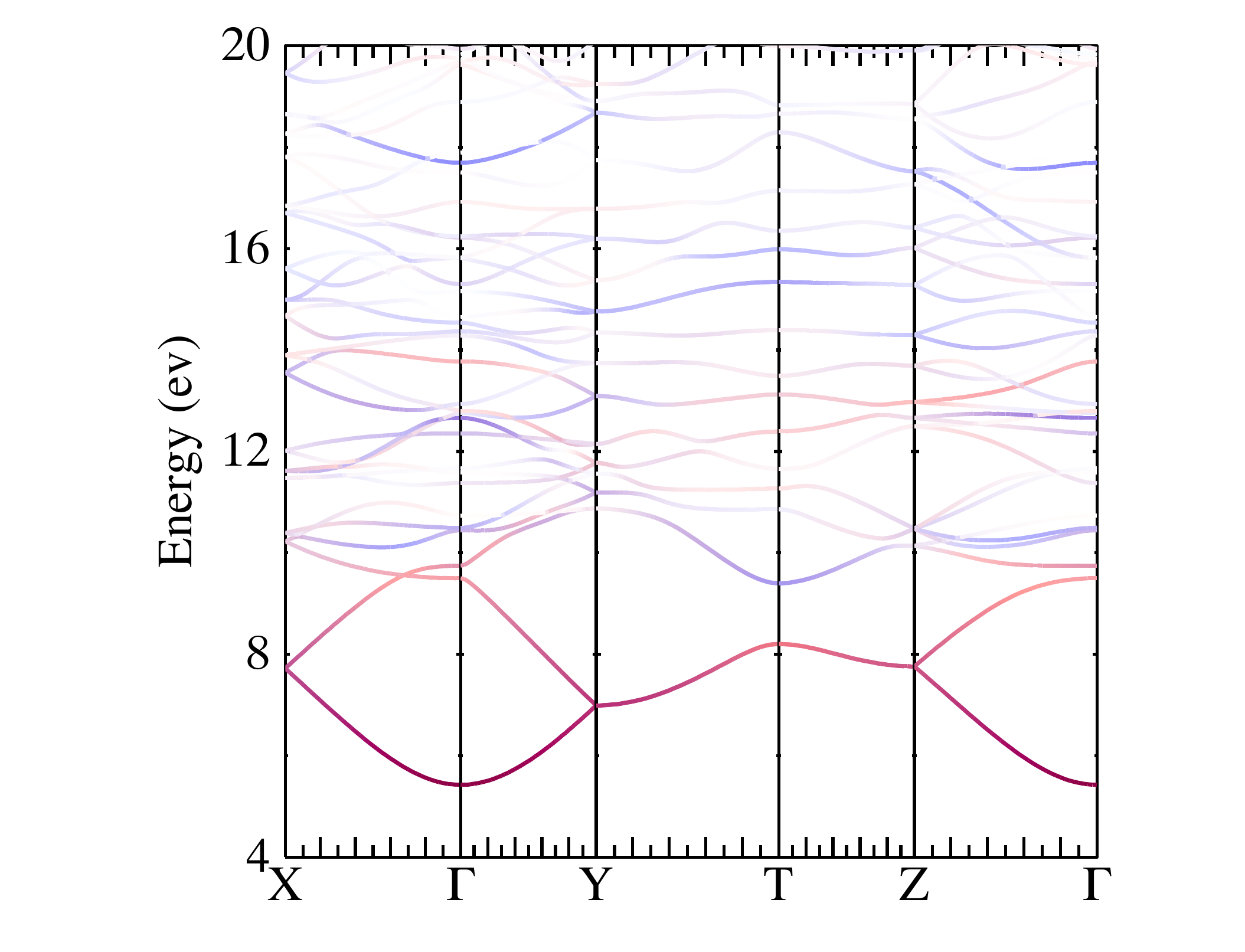}
  \caption{Conduction band structure of NaGaO$_2$ in $Pna2_1$ structure and $0.8\Sigma$
    approximation, showing Ga-$s$ and Na-$s$ contributions in red and blue
    respectively. \label{figbndcolngo}}
\end{figure}

The large energy scale band structure of $PNa2_1$ NaGaO$_2$
is shown in Fig. \ref{figbandngo}.
The partial densities of states in the valence and conduction band region
are given in Fig. \ref{figpdosngo}. They show similar to LiGaO$_2$
that the Ga-$3d$  bands lie above the O-$2s$ ones and that the Na contribution
to the conduction band PDOS occurs mainly well above the conduction band
minimum, while the latter is dominated by Ga-$4s$. This is also
shown in Fig. \ref{figbndcolngo} which shows the Ga-$4s$ and Na-$3s$
contributions to the conduction bands. 
In addition,
the Na-$2p$ semicore levels are seen to lie at about $-25$ eV.

\begin{figure}
    \includegraphics[width=0.5\textwidth]{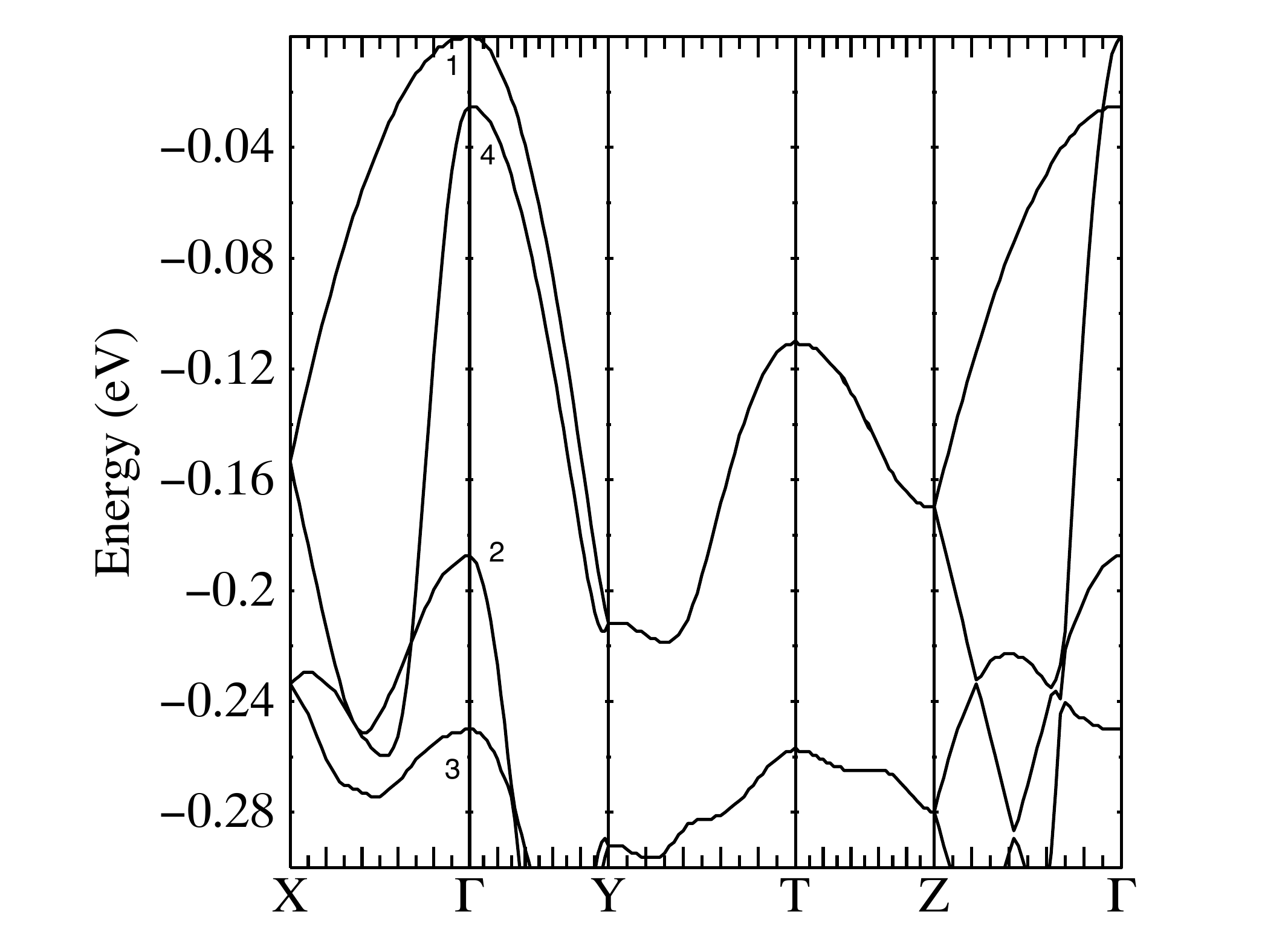}
    \includegraphics[width=0.5\textwidth]{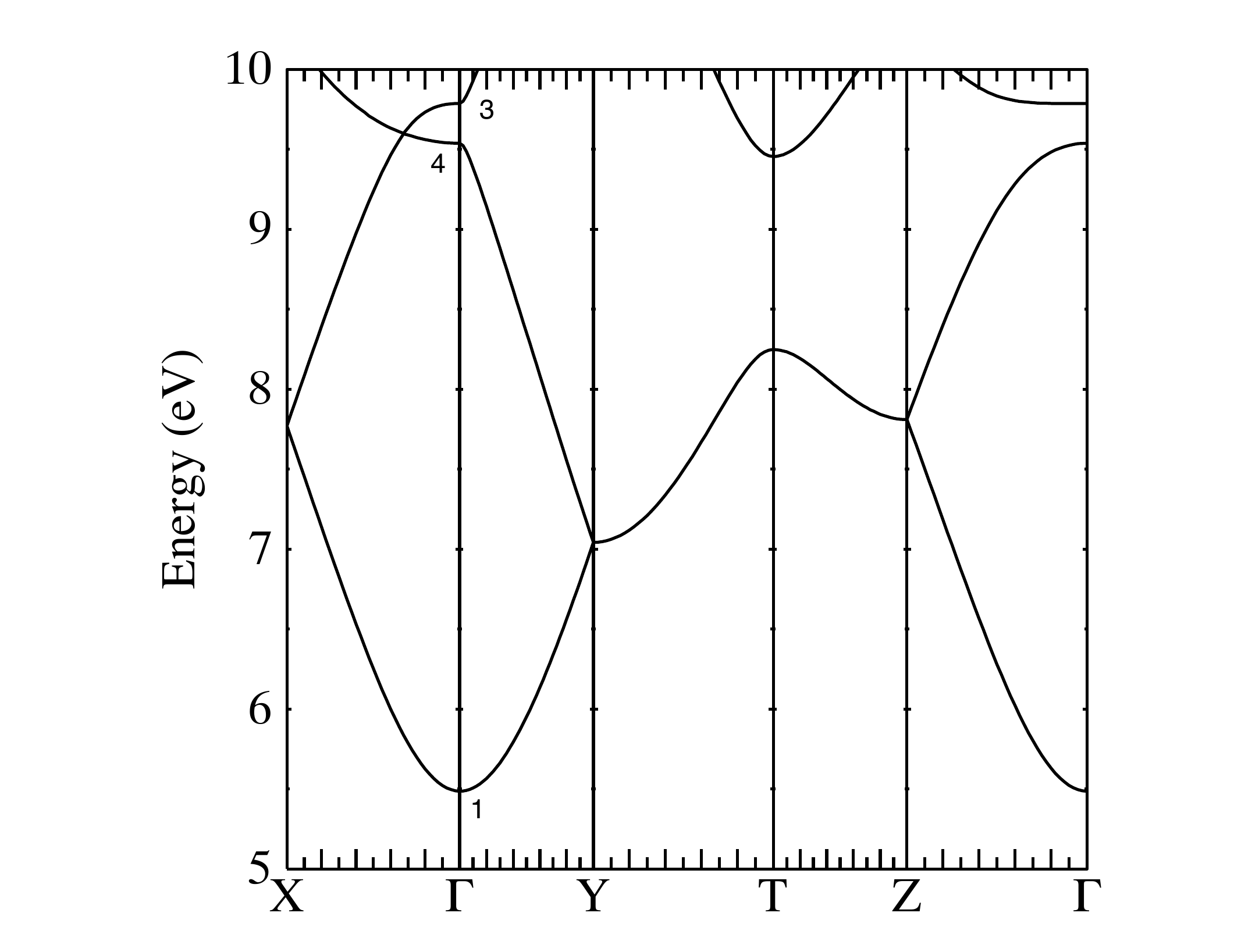}
    \caption{Symmetry labeled bands for NaGaO$_2$ in $Pna2_1$ structure.
      Please note different
      scales in (top) near valence band maximum) and (bottom)
      conduction band region.\label{figbndzoomngo}}
\end{figure}

\begin{table}
  \caption{Effective masses (in units of the free electron mass) and energy levels in NaGaO$_2$ in $Pna2_1$ structure.\label{tabmassngo}}
  \begin{ruledtabular}
    \begin{tabular}{lccccc}
 band& irrep& $E$ (eV) & $m_x$ & $m_y$ & $m_z$ \\ \hline
 CBM  &  $\Gamma_1$ & 5.486   & 0.33 & 0.35 & 0.35  \\ 
 VBM1 & $\Gamma_1$  & 0         & 6.8 & 2.7 & 0.5 \\
 VBM2 & $\Gamma_4$ & $-0.026$ & 0.5 & 2.7 & 9.5 \\
 VBM3 & $\Gamma_2$ & $-0.185$ & 3.4 & 0.6 & 3.1
    \end{tabular}
  \end{ruledtabular}
\end{table}

A zoom in on the valence band maximum and conduction band minimum
range are shown in Fig. \ref{figbndzoomngo}. 
The band splittings of the VBM and corresponding effective mass tensor
components are given in Table \ref{tabmassngo}.
The band gap is only slightly lower than for LiGaO$_2$ with
a value of 5.49 eV in the 0.8$\Sigma$ approximation and 2.88 eV in
the GGA approximation. The shift of the individual band edges between
GGA and 0.8$\Sigma$ $GW$ is $-1.49$ eV in the valence band and 1.12 eV
in the conduction band.  The band gap is probably still
slightly overestimated because of the GGA overestimate of the lattice constant.
Including a zero-point motion correction similar to LiGaO$_2$
and a lowering of gap by the deformation potential correction,
we estimate that the gap is 5.1$\pm0.1$ eV, which is only slightly lower
than in LiGaO$_2$ and still significantly higher than in $\beta$-Ga$_2$O$_3$.

\begin{figure}
  \includegraphics[width=9cm]{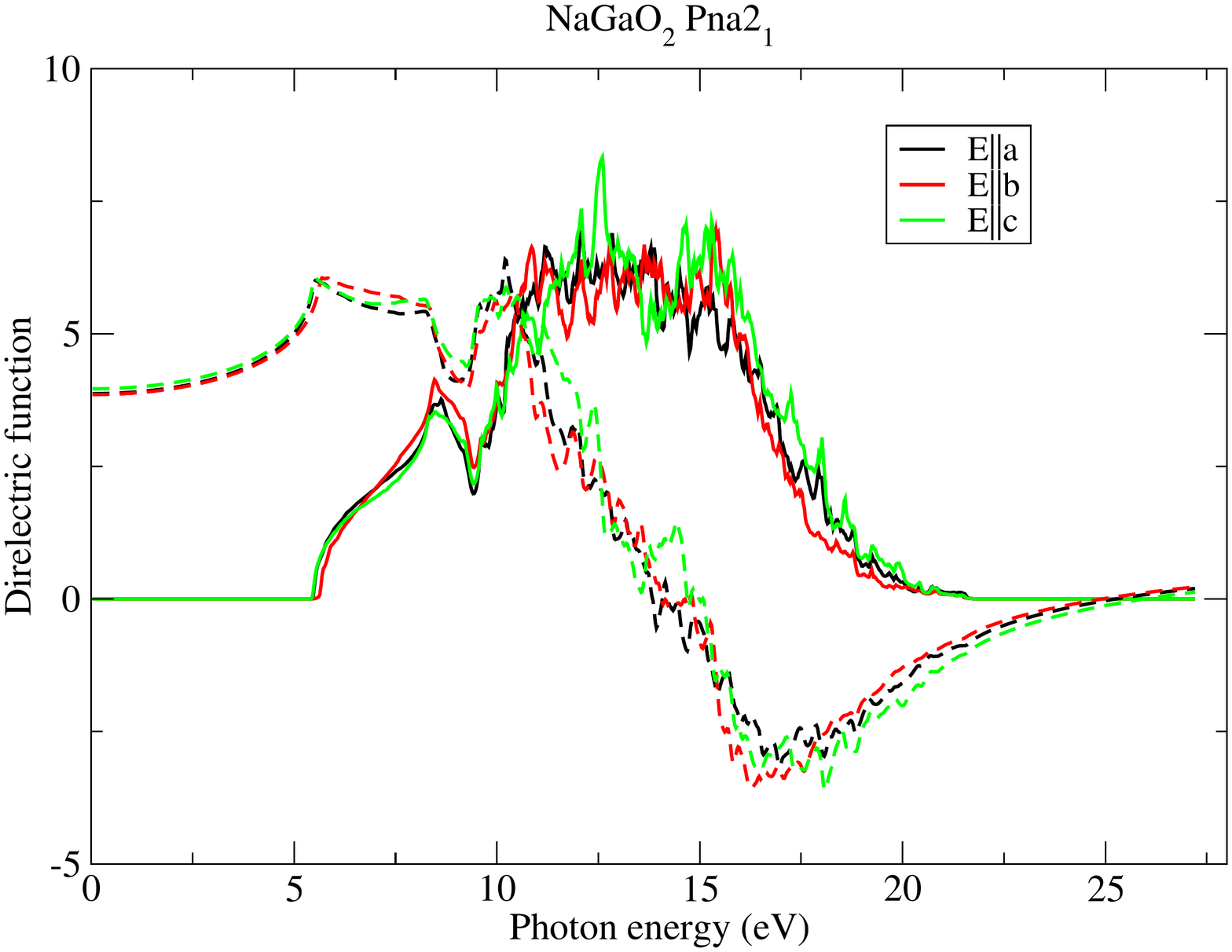}
  \caption{Imaginary $\varepsilon_2$  and real $\varepsilon_1$ parts
    of the optical dielectric function for $x,y,z$ directions for NaGaO$_2$
    in $Pna2_1$ structure.\label{figepsngo}}
\end{figure}

\begin{figure}
  \includegraphics[width=9cm]{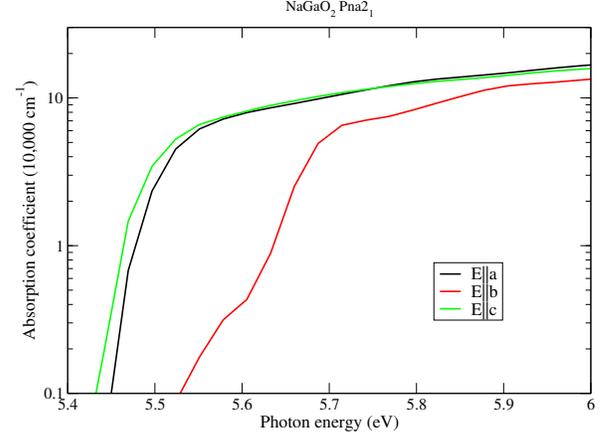}
  \caption{Absorption coefficient of NaGaO$_2$ in $Pna2_1$ structure.\label{figabsngo}}
\end{figure}

The dielectric functions and optical absorption
are given in Figs. \ref{figepsngo},\ref{figabsngo}.
They confirm the analysis of the optical anisotropy of the absorption onset
based on the symmetry labeled valence bands.  An interesting difference
from LiGaO$_2$ is that in NaGaO$_2$, the $x$  and $z$ polarization
onsets of absorption are close to each other while the $y$ onset is larger.
In contrast in LiGaO$_2$ the $x$ and $y$ onsets are close but both larger
than the $z$ onset. Thus for light incident on the basal plane (the $c$-plane)
there will be a larger anisotropy in the plane between the two
polarizations $x$ and $y$ for NaGaO$_2$ than for LiGaO$_2$.
On the other hand, in terms of transparent
conductor applications, we again see a large splitting between
the lowest and next higher conduction band of 4.07 eV, even larger than
in LiGaO$_2$.

The dielectric constant  $\varepsilon^\infty$  (diagonal) tensor
components and directional average are given in Table \ref{tabdielconstngo}
in different approximations. The trends are similar to the LiGaO$_2$
case. This suggests that the QS$GW$ calculation including the $d\Sigma/dk$
contribution to the matrix elements is an overestimate, perhaps because
of neglecting local field effects. The  renormalization of the matrix
elements gives values close to the GGA and is likely a more realistic
estimate.

\begin{table}
  \caption{Dielectric constants $\varepsilon^\infty$ of NaGaO$_2$ in
    various approximations.\label{tabdielconstngo}}
  \begin{ruledtabular}
    \begin{tabular}{ldddd}
       method & \multicolumn{1}{c}{$\varepsilon^\infty_x$} &  \multicolumn{1}{c}{$\varepsilon^\infty_x$} &  \multicolumn{1}{c}{$\varepsilon^\infty_x$} & \multicolumn{1}{c}{$(\varepsilon^\infty_x\varepsilon^\infty_y\varepsilon^\infty_z)^{1/3}$} \\ \hline
      GGA                              & 2.616 & 2.606 & 2.587 & 2.603 \\
      0.8$\Sigma$ $d\Sigma/dk$         & 4.268 & 4.130 & 4.286 & 4.227 \\
      0.8$\Sigma$ $d\Sigma/dk=0$       & 1.655 & 1.644 & 1.648 & 1.649 \\
      0.8$\Sigma$ rescaling            & 2.156 & 2.138 & 2.137 & 2.144 \\
    \end{tabular}
  \end{ruledtabular}
\end{table}

\subsection{Band structure of NaGaO$_2$ in $R\bar{3}m$ and $P4/mmm$ structures.}
\begin{figure}
  \includegraphics[width=9cm]{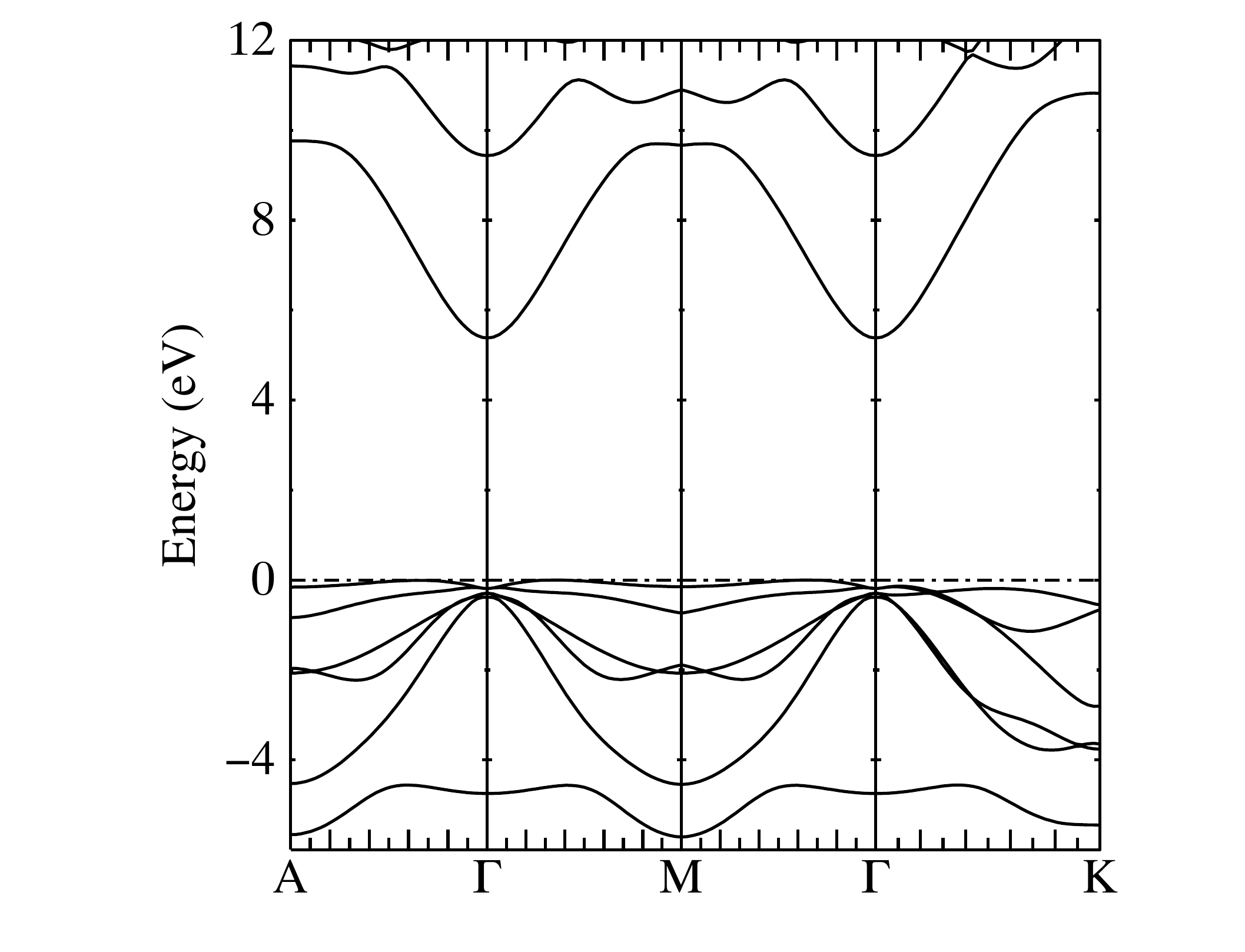}
  \includegraphics[width=9cm]{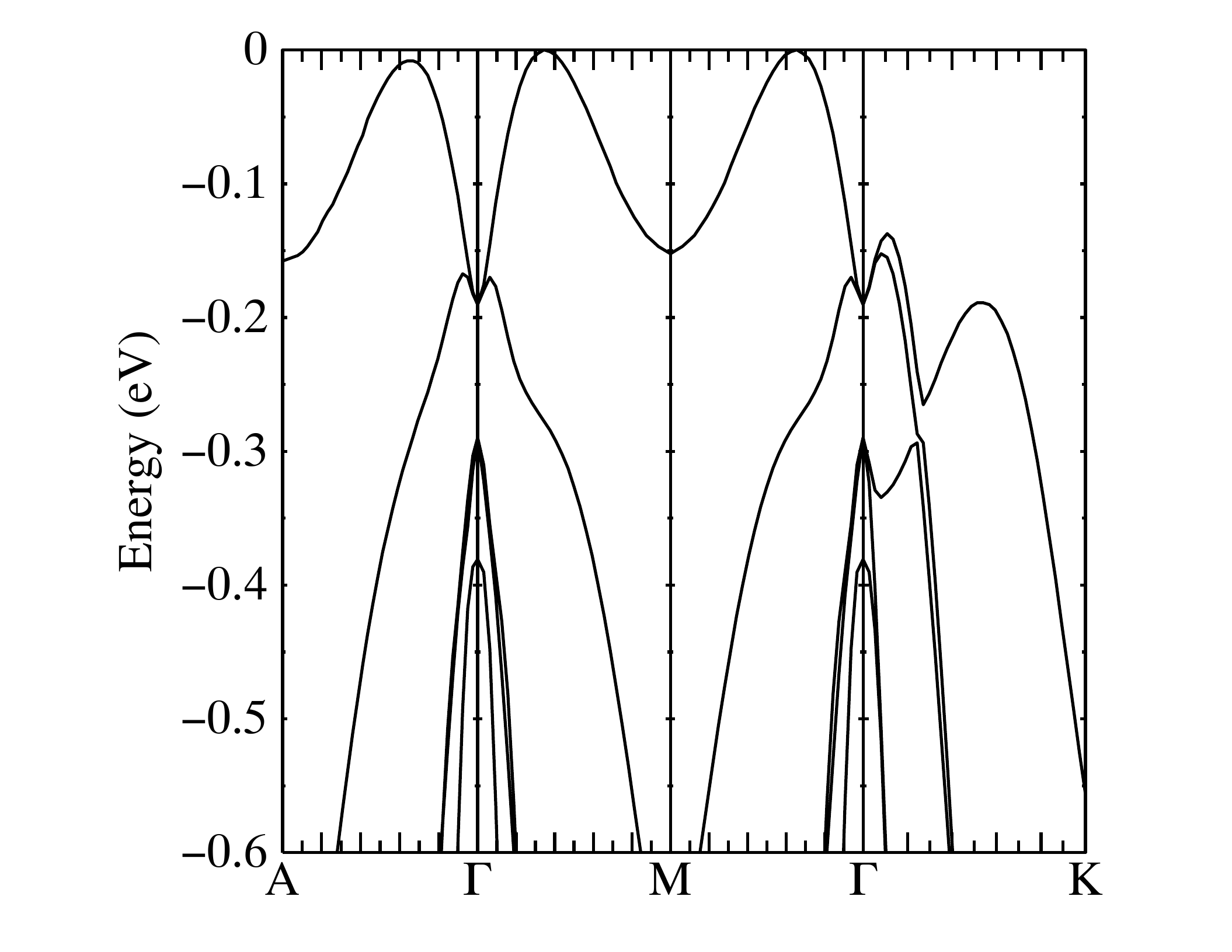}
  \caption{Band structure of NaGaO$_2$ in the $R\bar{3}m$ phase in 0.8$\Sigma$
    approximation. The bottom panel shows a zoom in on the VBM region.
    \label{figngor3m}}
\end{figure}

The band structure of NaGaO$_2$ in the $R\bar{3}m$ structure is shown
in Fig. \ref{figngor3m}. The band gap is again indirect and equal to
5.38 eV while the direct gap at $\Gamma$ is 5.57 eV.
On the other hand, in the $P4/mmm$ structure the gap is much smaller.
In fact, in GGA, there is a band overlap and in the 0.8$\Sigma$ QSGW approximation, the gap is only 0.965 eV. Interestingly, in this case the Ga-$3d$ and O-$2s$
band hybridize. 

\begin{figure}
  \includegraphics[width=9cm]{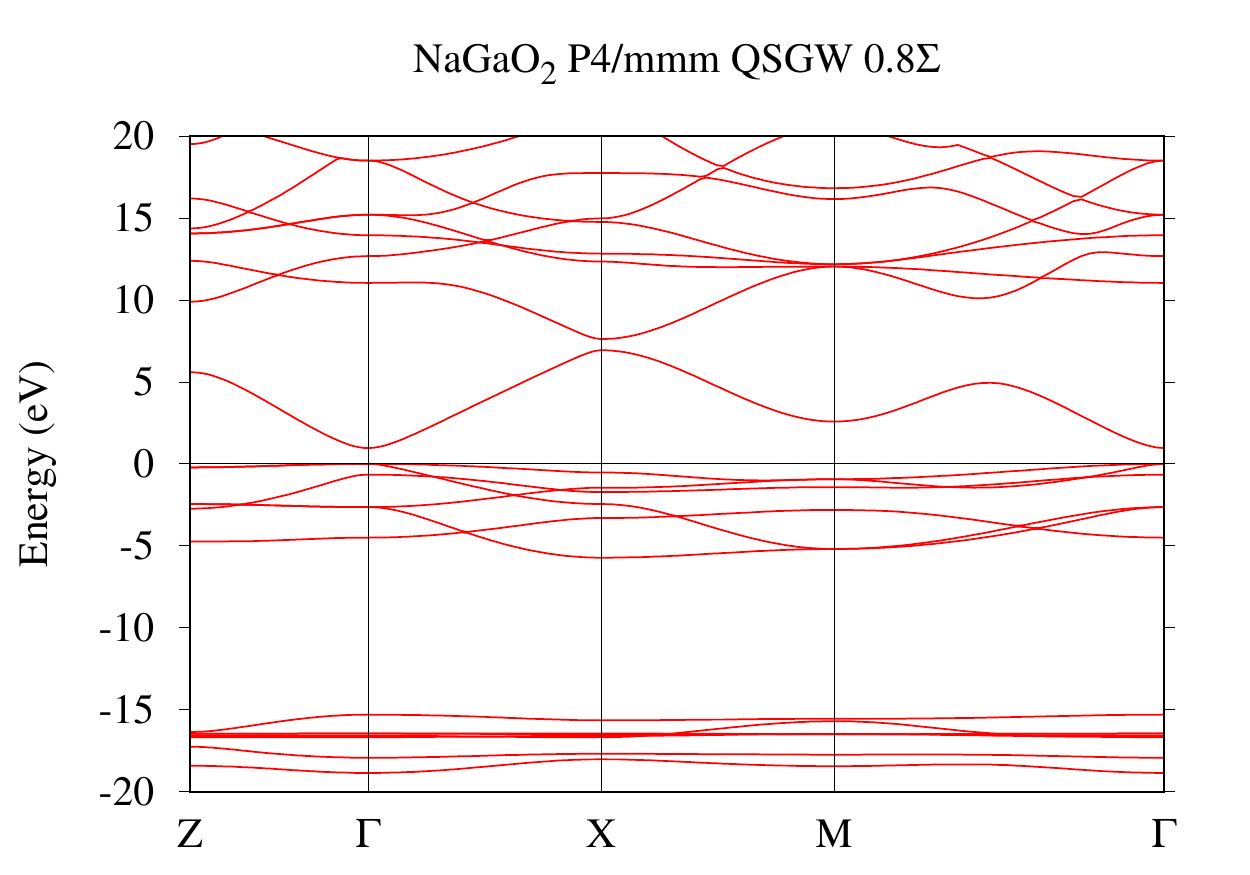}
  \caption{Band structure of NaGaO$_2$ in the $P4/mmm$ phase in 0.8$\Sigma$
    approximation.
    \label{figngop4}}
\end{figure}

\section{Conclusions}
In this paper we studied the band structures of LiGaO$_2$ and NaGaO$_2$
in three different crystal structures, the ambient pressure equilibrium
tetrahedrally bonded $Pna2_1$ and the high-pressure octahedral $R\bar{3}m$
and hypothetical $P4/mmm$  rocksalt type phase, all 
using the
QS$GW$ method. The $Pna2_1$ tetrahedrally bonded structure, which is
a cation ordered supercell of the parent wurtzite structure is found
in both cases to have lower energy than the octahedrally coordinated
$R\bar{3}m$ structure, which is found to be a high-pressure phase with
about 20 \% smaller volume per formula unit.  The materials
are ultra-wide band gap semiconductors with gaps of 5.8 eV (LiGaO$_2$)
and 5.5 eV (NaGaO$_2$)  in the $Pna2_1$ structure,
not including zero-point motion corrections, which are estimated
to be of order $-0.2$ eV. The gap in the high-pressure phase
are of similar magnitude but slightly indirect.  The valence band
is split in three levels due to the orthorhombic crystal field
splitting and lead to anisotropy of the optical absorption onset.
Effective mass tensors of the top three valence bands and the
conduction band were calculated.  In view of previous work, indicating
that LiGaO$_2$ can be n-type doped by Si or Ge, we consider
both materials to be promising as ultrawide gap semiconductors
for transparent conductor and high-power transistors. In particular,
the conduction band symmetry labeling  indicates that no optical
transitions can occur between the bottom of the conduction band
(when n-type doped) to higher conduction bands for energies less
than $\sim$4.0 eV. As part of this study, we also determined the transition
pressures from the tetrahedrally coordinated  $Pna2_1$ to the octahedrally
coordinated $R\bar{3}m$ phase and a rocksalt type $P4/mmm$
phase and found these to agree well with experiments for LiGaO$_2$. 

\bibliography{ligao2,ga2o3,dft,abinit,lmto,gw}

\begin{thebibliography}{44}%
\makeatletter
\providecommand \@ifxundefined [1]{%
 \@ifx{#1\undefined}
}%
\providecommand \@ifnum [1]{%
 \ifnum #1\expandafter \@firstoftwo
 \else \expandafter \@secondoftwo
 \fi
}%
\providecommand \@ifx [1]{%
 \ifx #1\expandafter \@firstoftwo
 \else \expandafter \@secondoftwo
 \fi
}%
\providecommand \natexlab [1]{#1}%
\providecommand \enquote  [1]{``#1''}%
\providecommand \bibnamefont  [1]{#1}%
\providecommand \bibfnamefont [1]{#1}%
\providecommand \citenamefont [1]{#1}%
\providecommand \href@noop [0]{\@secondoftwo}%
\providecommand \href [0]{\begingroup \@sanitize@url \@href}%
\providecommand \@href[1]{\@@startlink{#1}\@@href}%
\providecommand \@@href[1]{\endgroup#1\@@endlink}%
\providecommand \@sanitize@url [0]{\catcode `\\12\catcode `\$12\catcode
  `\&12\catcode `\#12\catcode `\^12\catcode `\_12\catcode `\%12\relax}%
\providecommand \@@startlink[1]{}%
\providecommand \@@endlink[0]{}%
\providecommand \url  [0]{\begingroup\@sanitize@url \@url }%
\providecommand \@url [1]{\endgroup\@href {#1}{\urlprefix }}%
\providecommand \urlprefix  [0]{URL }%
\providecommand \Eprint [0]{\href }%
\providecommand \doibase [0]{http://dx.doi.org/}%
\providecommand \selectlanguage [0]{\@gobble}%
\providecommand \bibinfo  [0]{\@secondoftwo}%
\providecommand \bibfield  [0]{\@secondoftwo}%
\providecommand \translation [1]{[#1]}%
\providecommand \BibitemOpen [0]{}%
\providecommand \bibitemStop [0]{}%
\providecommand \bibitemNoStop [0]{.\EOS\space}%
\providecommand \EOS [0]{\spacefactor3000\relax}%
\providecommand \BibitemShut  [1]{\csname bibitem#1\endcsname}%
\let\auto@bib@innerbib\@empty
%</preamble>
\bibitem [{\citenamefont {Nanamatsu}\ \emph {et~al.}(1972)\citenamefont
  {Nanamatsu}, \citenamefont {Doi},\ and\ \citenamefont
  {Takahashi}}]{Nanamatsu72}%
  \BibitemOpen
  \bibfield  {author} {\bibinfo {author} {\bibfnamefont {S.}~\bibnamefont
  {Nanamatsu}}, \bibinfo {author} {\bibfnamefont {K.}~\bibnamefont {Doi}}, \
  and\ \bibinfo {author} {\bibfnamefont {M.}~\bibnamefont {Takahashi}},\ }\href
  {\doibase 10.1143/jjap.11.816} {\bibfield  {journal} {\bibinfo  {journal}
  {Japanese Journal of Applied Physics}\ }\textbf {\bibinfo {volume} {11}},\
  \bibinfo {pages} {816} (\bibinfo {year} {1972})}\BibitemShut {NoStop}%
\bibitem [{\citenamefont {Gupta}\ \emph {et~al.}(1976)\citenamefont {Gupta},
  \citenamefont {Vetelino}, \citenamefont {Jipson},\ and\ \citenamefont
  {Field}}]{Gupta76}%
  \BibitemOpen
  \bibfield  {author} {\bibinfo {author} {\bibfnamefont {S.~N.}\ \bibnamefont
  {Gupta}}, \bibinfo {author} {\bibfnamefont {J.~F.}\ \bibnamefont {Vetelino}},
  \bibinfo {author} {\bibfnamefont {V.~B.}\ \bibnamefont {Jipson}}, \ and\
  \bibinfo {author} {\bibfnamefont {J.~C.}\ \bibnamefont {Field}},\ }\href
  {\doibase 10.1063/1.322719} {\bibfield  {journal} {\bibinfo  {journal}
  {Journal of Applied Physics}\ }\textbf {\bibinfo {volume} {47}},\ \bibinfo
  {pages} {858} (\bibinfo {year} {1976})}\BibitemShut {NoStop}%
\bibitem [{\citenamefont {Boonchun}\ and\ \citenamefont
  {Lambrecht}(2010)}]{Boonchun10}%
  \BibitemOpen
  \bibfield  {author} {\bibinfo {author} {\bibfnamefont {A.}~\bibnamefont
  {Boonchun}}\ and\ \bibinfo {author} {\bibfnamefont {W.~R.~L.}\ \bibnamefont
  {Lambrecht}},\ }\href {\doibase 10.1103/PhysRevB.81.235214} {\bibfield
  {journal} {\bibinfo  {journal} {Phys. Rev. B}\ }\textbf {\bibinfo {volume}
  {81}},\ \bibinfo {pages} {235214} (\bibinfo {year} {2010})}\BibitemShut
  {NoStop}%
\bibitem [{\citenamefont {Rashkeev}\ \emph {et~al.}(1999)\citenamefont
  {Rashkeev}, \citenamefont {Limpijumnong},\ and\ \citenamefont
  {Lambrecht}}]{Rashkeev99}%
  \BibitemOpen
  \bibfield  {author} {\bibinfo {author} {\bibfnamefont {S.~N.}\ \bibnamefont
  {Rashkeev}}, \bibinfo {author} {\bibfnamefont {S.}~\bibnamefont
  {Limpijumnong}}, \ and\ \bibinfo {author} {\bibfnamefont {W.~R.~L.}\
  \bibnamefont {Lambrecht}},\ }\href {\doibase 10.1364/JOSAB.16.002217}
  {\bibfield  {journal} {\bibinfo  {journal} {J. Opt. Soc. Am. B}\ }\textbf
  {\bibinfo {volume} {16}},\ \bibinfo {pages} {2217} (\bibinfo {year}
  {1999})}\BibitemShut {NoStop}%
\bibitem [{\citenamefont {Marezio}(1965)}]{Marezio65}%
  \BibitemOpen
  \bibfield  {author} {\bibinfo {author} {\bibfnamefont {M.}~\bibnamefont
  {Marezio}},\ }\href {\doibase 10.1107/S0365110X65001068} {\bibfield
  {journal} {\bibinfo  {journal} {Acta Crystallographica}\ }\textbf {\bibinfo
  {volume} {18}},\ \bibinfo {pages} {481} (\bibinfo {year} {1965})}\BibitemShut
  {NoStop}%
\bibitem [{\citenamefont {Ishii}\ \emph {et~al.}(1998)\citenamefont {Ishii},
  \citenamefont {Tazoh},\ and\ \citenamefont {Miyazawa}}]{Ishii98}%
  \BibitemOpen
  \bibfield  {author} {\bibinfo {author} {\bibfnamefont {T.}~\bibnamefont
  {Ishii}}, \bibinfo {author} {\bibfnamefont {Y.}~\bibnamefont {Tazoh}}, \ and\
  \bibinfo {author} {\bibfnamefont {S.}~\bibnamefont {Miyazawa}},\ }\href
  {\doibase https://doi.org/10.1016/S0022-0248(97)00510-1} {\bibfield
  {journal} {\bibinfo  {journal} {J. Crystal Growth}\ }\textbf {\bibinfo
  {volume} {186}},\ \bibinfo {pages} {409 } (\bibinfo {year}
  {1998})}\BibitemShut {NoStop}%
\bibitem [{\citenamefont {Sakurada}\ \emph {et~al.}(2007)\citenamefont
  {Sakurada}, \citenamefont {Kobayashi}, \citenamefont {Kawaguchi},
  \citenamefont {Ohta},\ and\ \citenamefont {Fujioka}}]{Sakurada07}%
  \BibitemOpen
  \bibfield  {author} {\bibinfo {author} {\bibfnamefont {K.}~\bibnamefont
  {Sakurada}}, \bibinfo {author} {\bibfnamefont {A.}~\bibnamefont {Kobayashi}},
  \bibinfo {author} {\bibfnamefont {Y.}~\bibnamefont {Kawaguchi}}, \bibinfo
  {author} {\bibfnamefont {J.}~\bibnamefont {Ohta}}, \ and\ \bibinfo {author}
  {\bibfnamefont {H.}~\bibnamefont {Fujioka}},\ }\href {\doibase
  10.1063/1.2737928} {\bibfield  {journal} {\bibinfo  {journal} {Applied
  Physics Letters}\ }\textbf {\bibinfo {volume} {90}},\ \bibinfo {pages}
  {211913} (\bibinfo {year} {2007})},\ \Eprint
  {http://arxiv.org/abs/https://doi.org/10.1063/1.2737928}
  {https://doi.org/10.1063/1.2737928} \BibitemShut {NoStop}%
\bibitem [{\citenamefont {Omata}\ \emph {et~al.}(2008)\citenamefont {Omata},
  \citenamefont {Tanaka}, \citenamefont {Tazuke}, \citenamefont {Nose},\ and\
  \citenamefont {Otsuka-Yao-Matsuo}}]{Omata08}%
  \BibitemOpen
  \bibfield  {author} {\bibinfo {author} {\bibfnamefont {T.}~\bibnamefont
  {Omata}}, \bibinfo {author} {\bibfnamefont {K.}~\bibnamefont {Tanaka}},
  \bibinfo {author} {\bibfnamefont {A.}~\bibnamefont {Tazuke}}, \bibinfo
  {author} {\bibfnamefont {K.}~\bibnamefont {Nose}}, \ and\ \bibinfo {author}
  {\bibfnamefont {S.}~\bibnamefont {Otsuka-Yao-Matsuo}},\ }\href {\doibase
  10.1063/1.2903906} {\bibfield  {journal} {\bibinfo  {journal} {J. Appl.
  Phys.}\ }\textbf {\bibinfo {volume} {103}},\ \bibinfo {pages} {083706}
  (\bibinfo {year} {2008})}\BibitemShut {NoStop}%
\bibitem [{\citenamefont {Omata}\ \emph {et~al.}(2011)\citenamefont {Omata},
  \citenamefont {Kita}, \citenamefont {Nose}, \citenamefont {Tachibana},\ and\
  \citenamefont {Otsuka-Yao-Matsuo}}]{Omata11}%
  \BibitemOpen
  \bibfield  {author} {\bibinfo {author} {\bibfnamefont {T.}~\bibnamefont
  {Omata}}, \bibinfo {author} {\bibfnamefont {M.}~\bibnamefont {Kita}},
  \bibinfo {author} {\bibfnamefont {K.}~\bibnamefont {Nose}}, \bibinfo {author}
  {\bibfnamefont {K.}~\bibnamefont {Tachibana}}, \ and\ \bibinfo {author}
  {\bibfnamefont {S.}~\bibnamefont {Otsuka-Yao-Matsuo}},\ }\href {\doibase
  10.1143/jjap.50.031102} {\bibfield  {journal} {\bibinfo  {journal} {Jpn. J.
  Appl. Phys.}\ }\textbf {\bibinfo {volume} {50}},\ \bibinfo {pages} {031102}
  (\bibinfo {year} {2011})}\BibitemShut {NoStop}%
\bibitem [{\citenamefont {Ohkubo}\ \emph {et~al.}(2002)\citenamefont {Ohkubo},
  \citenamefont {Hirose}, \citenamefont {Tamura}, \citenamefont {Nishii},
  \citenamefont {Saito}, \citenamefont {Koinuma}, \citenamefont {Ahemt},
  \citenamefont {Chikyow}, \citenamefont {Ishii}, \citenamefont {Miyazawa},
  \citenamefont {Segawa}, \citenamefont {Fukumura},\ and\ \citenamefont
  {Kawasaki}}]{Ohkubo2002}%
  \BibitemOpen
  \bibfield  {author} {\bibinfo {author} {\bibfnamefont {I.}~\bibnamefont
  {Ohkubo}}, \bibinfo {author} {\bibfnamefont {C.}~\bibnamefont {Hirose}},
  \bibinfo {author} {\bibfnamefont {K.}~\bibnamefont {Tamura}}, \bibinfo
  {author} {\bibfnamefont {J.}~\bibnamefont {Nishii}}, \bibinfo {author}
  {\bibfnamefont {H.}~\bibnamefont {Saito}}, \bibinfo {author} {\bibfnamefont
  {H.}~\bibnamefont {Koinuma}}, \bibinfo {author} {\bibfnamefont
  {P.}~\bibnamefont {Ahemt}}, \bibinfo {author} {\bibfnamefont
  {T.}~\bibnamefont {Chikyow}}, \bibinfo {author} {\bibfnamefont
  {T.}~\bibnamefont {Ishii}}, \bibinfo {author} {\bibfnamefont
  {S.}~\bibnamefont {Miyazawa}}, \bibinfo {author} {\bibfnamefont
  {Y.}~\bibnamefont {Segawa}}, \bibinfo {author} {\bibfnamefont
  {T.}~\bibnamefont {Fukumura}}, \ and\ \bibinfo {author} {\bibfnamefont
  {M.}~\bibnamefont {Kawasaki}},\ }\href {\doibase 10.1063/1.1512311}
  {\bibfield  {journal} {\bibinfo  {journal} {Journal of Applied Physics}\
  }\textbf {\bibinfo {volume} {92}},\ \bibinfo {pages} {5587} (\bibinfo {year}
  {2002})}\BibitemShut {NoStop}%
\bibitem [{\citenamefont {Boonchun}\ and\ \citenamefont
  {Lambrecht}(2011)}]{Boonchun11SPIE}%
  \BibitemOpen
  \bibfield  {author} {\bibinfo {author} {\bibfnamefont {A.}~\bibnamefont
  {Boonchun}}\ and\ \bibinfo {author} {\bibfnamefont {W.~R.~L.}\ \bibnamefont
  {Lambrecht}},\ }in\ \href {\doibase 10.1117/12.879320} {\emph {\bibinfo
  {booktitle} {{Oxide-based Materials and Devices II}}}},\ \bibinfo {series}
  {Proceedings of SPIE}, Vol.\ \bibinfo {volume} {7940},\ \bibinfo {editor}
  {edited by\ \bibinfo {editor} {\bibfnamefont {F.~H.}\ \bibnamefont {Terani}},
  \bibinfo {editor} {\bibfnamefont {D.~C.}\ \bibnamefont {Look}}, \ and\
  \bibinfo {editor} {\bibfnamefont {D.~J.}\ \bibnamefont {Rogers}}},\ \bibinfo
  {organization} {International Society for Optics and Photonics}\ (\bibinfo
  {publisher} {SPIE},\ \bibinfo {year} {2011})\ pp.\ \bibinfo {pages}
  {129--134}\BibitemShut {NoStop}%
\bibitem [{\citenamefont {Boonchun}\ \emph {et~al.}(2019)\citenamefont
  {Boonchun}, \citenamefont {Dabsamut},\ and\ \citenamefont
  {Lambrecht}}]{Boonchun19}%
  \BibitemOpen
  \bibfield  {author} {\bibinfo {author} {\bibfnamefont {A.}~\bibnamefont
  {Boonchun}}, \bibinfo {author} {\bibfnamefont {K.}~\bibnamefont {Dabsamut}},
  \ and\ \bibinfo {author} {\bibfnamefont {W.~R.~L.}\ \bibnamefont
  {Lambrecht}},\ }\href {\doibase 10.1063/1.5126028} {\bibfield  {journal}
  {\bibinfo  {journal} {Journal of Applied Physics}\ }\textbf {\bibinfo
  {volume} {126}},\ \bibinfo {pages} {155703} (\bibinfo {year}
  {2019})}\BibitemShut {NoStop}%
\bibitem [{\citenamefont {Dabsamut}\ \emph {et~al.}(2020)\citenamefont
  {Dabsamut}, \citenamefont {Boonchun},\ and\ \citenamefont
  {Lambrecht}}]{Boonchun20}%
  \BibitemOpen
  \bibfield  {author} {\bibinfo {author} {\bibfnamefont {K.}~\bibnamefont
  {Dabsamut}}, \bibinfo {author} {\bibfnamefont {A.}~\bibnamefont {Boonchun}},
  \ and\ \bibinfo {author} {\bibfnamefont {W.~R.~L.}\ \bibnamefont
  {Lambrecht}},\ }\href {http://iopscience.iop.org/10.1088/1361-6463/ab8514}
  {\bibfield  {journal} {\bibinfo  {journal} {Journal of Physics D: Applied
  Physics}\ } (\bibinfo {year} {2020})}\BibitemShut {NoStop}%
\bibitem [{\citenamefont {Lenyk}\ \emph {et~al.}(2018)\citenamefont {Lenyk},
  \citenamefont {Holston}, \citenamefont {Kananen}, \citenamefont
  {Halliburton},\ and\ \citenamefont {Giles}}]{Lenyk18}%
  \BibitemOpen
  \bibfield  {author} {\bibinfo {author} {\bibfnamefont {C.~A.}\ \bibnamefont
  {Lenyk}}, \bibinfo {author} {\bibfnamefont {M.~S.}\ \bibnamefont {Holston}},
  \bibinfo {author} {\bibfnamefont {B.~E.}\ \bibnamefont {Kananen}}, \bibinfo
  {author} {\bibfnamefont {L.~E.}\ \bibnamefont {Halliburton}}, \ and\ \bibinfo
  {author} {\bibfnamefont {N.~C.}\ \bibnamefont {Giles}},\ }\href {\doibase
  10.1063/1.5050532} {\bibfield  {journal} {\bibinfo  {journal} {Journal of
  Applied Physics}\ }\textbf {\bibinfo {volume} {124}},\ \bibinfo {pages}
  {135702} (\bibinfo {year} {2018})}\BibitemShut {NoStop}%
\bibitem [{\citenamefont {Skachkov}\ \emph {et~al.}(2020)\citenamefont
  {Skachkov}, \citenamefont {Lambrecht}, \citenamefont {Dabsamut},\ and\
  \citenamefont {Boonchun}}]{Skachkov20}%
  \BibitemOpen
  \bibfield  {author} {\bibinfo {author} {\bibfnamefont {D.}~\bibnamefont
  {Skachkov}}, \bibinfo {author} {\bibfnamefont {W.~R.~L.}\ \bibnamefont
  {Lambrecht}}, \bibinfo {author} {\bibfnamefont {K.}~\bibnamefont {Dabsamut}},
  \ and\ \bibinfo {author} {\bibfnamefont {A.}~\bibnamefont {Boonchun}},\
  }\href {\doibase 10.1088/1361-6463/ab6f1c} {\bibfield  {journal} {\bibinfo
  {journal} {Journal of Physics D: Applied Physics}\ }\textbf {\bibinfo
  {volume} {53}},\ \bibinfo {pages} {17LT01} (\bibinfo {year}
  {2020})}\BibitemShut {NoStop}%
\bibitem [{\citenamefont {Sasaki}\ \emph {et~al.}(2013)\citenamefont {Sasaki},
  \citenamefont {Higashiwaki}, \citenamefont {Kuramata}, \citenamefont
  {Masui},\ and\ \citenamefont {Yamakoshi}}]{Sasaki13}%
  \BibitemOpen
  \bibfield  {author} {\bibinfo {author} {\bibfnamefont {K.}~\bibnamefont
  {Sasaki}}, \bibinfo {author} {\bibfnamefont {M.}~\bibnamefont {Higashiwaki}},
  \bibinfo {author} {\bibfnamefont {A.}~\bibnamefont {Kuramata}}, \bibinfo
  {author} {\bibfnamefont {T.}~\bibnamefont {Masui}}, \ and\ \bibinfo {author}
  {\bibfnamefont {S.}~\bibnamefont {Yamakoshi}},\ }\href {\doibase
  http://dx.doi.org/10.1016/j.jcrysgro.2013.02.015} {\bibfield  {journal}
  {\bibinfo  {journal} {J. Cryst. Growth}\ }\textbf {\bibinfo {volume} {378}},\
  \bibinfo {pages} {591 } (\bibinfo {year} {2013})},\ \bibinfo {note} {the 17th
  International Conference on Molecular Beam Epitaxy}\BibitemShut {NoStop}%
\bibitem [{\citenamefont {Green}\ \emph {et~al.}(2016)\citenamefont {Green},
  \citenamefont {Chabak}, \citenamefont {Heller}, \citenamefont {Fitch},
  \citenamefont {Baldini}, \citenamefont {Fiedler}, \citenamefont {Irmscher},
  \citenamefont {Wagner}, \citenamefont {Galazka}, \citenamefont {Tetlak},
  \citenamefont {Crespo}, \citenamefont {Leedy},\ and\ \citenamefont
  {Jessen}}]{Green16}%
  \BibitemOpen
  \bibfield  {author} {\bibinfo {author} {\bibfnamefont {A.~J.}\ \bibnamefont
  {Green}}, \bibinfo {author} {\bibfnamefont {K.~D.}\ \bibnamefont {Chabak}},
  \bibinfo {author} {\bibfnamefont {E.~R.}\ \bibnamefont {Heller}}, \bibinfo
  {author} {\bibfnamefont {R.~C.}\ \bibnamefont {Fitch}}, \bibinfo {author}
  {\bibfnamefont {M.}~\bibnamefont {Baldini}}, \bibinfo {author} {\bibfnamefont
  {A.}~\bibnamefont {Fiedler}}, \bibinfo {author} {\bibfnamefont
  {K.}~\bibnamefont {Irmscher}}, \bibinfo {author} {\bibfnamefont
  {G.}~\bibnamefont {Wagner}}, \bibinfo {author} {\bibfnamefont
  {Z.}~\bibnamefont {Galazka}}, \bibinfo {author} {\bibfnamefont {S.~E.}\
  \bibnamefont {Tetlak}}, \bibinfo {author} {\bibfnamefont {A.}~\bibnamefont
  {Crespo}}, \bibinfo {author} {\bibfnamefont {K.}~\bibnamefont {Leedy}}, \
  and\ \bibinfo {author} {\bibfnamefont {G.~H.}\ \bibnamefont {Jessen}},\
  }\href {\doibase 10.1109/LED.2016.2568139} {\bibfield  {journal} {\bibinfo
  {journal} {IEEE Electron Device Letters}\ }\textbf {\bibinfo {volume} {37}},\
  \bibinfo {pages} {902} (\bibinfo {year} {2016})}\BibitemShut {NoStop}%
\bibitem [{\citenamefont {Wolan}\ and\ \citenamefont
  {Hoflund}(1998)}]{Wolan98}%
  \BibitemOpen
  \bibfield  {author} {\bibinfo {author} {\bibfnamefont {J.~T.}\ \bibnamefont
  {Wolan}}\ and\ \bibinfo {author} {\bibfnamefont {G.~B.}\ \bibnamefont
  {Hoflund}},\ }\href {\doibase 10.1116/1.581495} {\bibfield  {journal}
  {\bibinfo  {journal} {J. Vac. Sci. Tech. A}\ }\textbf {\bibinfo {volume}
  {16}},\ \bibinfo {pages} {3414} (\bibinfo {year} {1998})}\BibitemShut
  {NoStop}%
\bibitem [{\citenamefont {Chen}\ \emph {et~al.}(2014)\citenamefont {Chen},
  \citenamefont {Li}, \citenamefont {Yu},\ and\ \citenamefont {Chou}}]{Chen14}%
  \BibitemOpen
  \bibfield  {author} {\bibinfo {author} {\bibfnamefont {C.}~\bibnamefont
  {Chen}}, \bibinfo {author} {\bibfnamefont {C.-A.}\ \bibnamefont {Li}},
  \bibinfo {author} {\bibfnamefont {S.-H.}\ \bibnamefont {Yu}}, \ and\ \bibinfo
  {author} {\bibfnamefont {M.~M.}\ \bibnamefont {Chou}},\ }\href {\doibase
  https://doi.org/10.1016/j.jcrysgro.2014.06.040} {\bibfield  {journal}
  {\bibinfo  {journal} {Journal of Crystal Growth}\ }\textbf {\bibinfo {volume}
  {402}},\ \bibinfo {pages} {325 } (\bibinfo {year} {2014})}\BibitemShut
  {NoStop}%
\bibitem [{\citenamefont {Johnson}\ \emph {et~al.}(2011)\citenamefont
  {Johnson}, \citenamefont {McLeod},\ and\ \citenamefont
  {Moewes}}]{Johnson2011}%
  \BibitemOpen
  \bibfield  {author} {\bibinfo {author} {\bibfnamefont {N.~W.}\ \bibnamefont
  {Johnson}}, \bibinfo {author} {\bibfnamefont {J.~A.}\ \bibnamefont {McLeod}},
  \ and\ \bibinfo {author} {\bibfnamefont {A.}~\bibnamefont {Moewes}},\ }\href
  {\doibase 10.1088/0953-8984/23/44/445501} {\bibfield  {journal} {\bibinfo
  {journal} {Journal of Physics: Condensed Matter}\ }\textbf {\bibinfo {volume}
  {23}},\ \bibinfo {pages} {445501} (\bibinfo {year} {2011})}\BibitemShut
  {NoStop}%
\bibitem [{\citenamefont {Ratnaparkhe}\ and\ \citenamefont
  {Lambrecht}(2020)}]{Amolalgo}%
  \BibitemOpen
  \bibfield  {author} {\bibinfo {author} {\bibfnamefont {A.}~\bibnamefont
  {Ratnaparkhe}}\ and\ \bibinfo {author} {\bibfnamefont {W.~R.~L.}\
  \bibnamefont {Lambrecht}},\ }\href {\doibase 10.1002/pssb.201900317}
  {\bibfield  {journal} {\bibinfo  {journal} {Physica Status Solidi (b)}\
  }\textbf {\bibinfo {volume} {257}},\ \bibinfo {pages} {1900317} (\bibinfo
  {year} {2020})}\BibitemShut {NoStop}%
\bibitem [{\citenamefont {Ratnaparkhe}\ and\ \citenamefont
  {Lambrecht}(2017)}]{Ratnaparkhe17}%
  \BibitemOpen
  \bibfield  {author} {\bibinfo {author} {\bibfnamefont {A.}~\bibnamefont
  {Ratnaparkhe}}\ and\ \bibinfo {author} {\bibfnamefont {W.~R.~L.}\
  \bibnamefont {Lambrecht}},\ }\href {\doibase 10.1063/1.4978668} {\bibfield
  {journal} {\bibinfo  {journal} {Applied Physics Letters}\ }\textbf {\bibinfo
  {volume} {110}},\ \bibinfo {pages} {132103} (\bibinfo {year}
  {2017})}\BibitemShut {NoStop}%
\bibitem [{MP()}]{MP}%
  \BibitemOpen
  MP,\ \href {https://materialsproject.org/} {}\bibinfo {note}
  {\url{https://materialsproject.org/}}\BibitemShut {NoStop}%
\bibitem [{que()}]{questaal}%
  \BibitemOpen
  \href@noop {} {}\bibinfo {howpublished}
  {\url{http://www.questaal.org/}}\BibitemShut {NoStop}%
\bibitem [{\citenamefont {Pashov}\ \emph {et~al.}(2019)\citenamefont {Pashov},
  \citenamefont {Acharya}, \citenamefont {Lambrecht}, \citenamefont {Jackson},
  \citenamefont {Belashchenko}, \citenamefont {Chantis}, \citenamefont
  {Jamet},\ and\ \citenamefont {van Schilfgaarde}}]{questaalpaper}%
  \BibitemOpen
  \bibfield  {author} {\bibinfo {author} {\bibfnamefont {D.}~\bibnamefont
  {Pashov}}, \bibinfo {author} {\bibfnamefont {S.}~\bibnamefont {Acharya}},
  \bibinfo {author} {\bibfnamefont {W.~R.}\ \bibnamefont {Lambrecht}}, \bibinfo
  {author} {\bibfnamefont {J.}~\bibnamefont {Jackson}}, \bibinfo {author}
  {\bibfnamefont {K.~D.}\ \bibnamefont {Belashchenko}}, \bibinfo {author}
  {\bibfnamefont {A.}~\bibnamefont {Chantis}}, \bibinfo {author} {\bibfnamefont
  {F.}~\bibnamefont {Jamet}}, \ and\ \bibinfo {author} {\bibfnamefont
  {M.}~\bibnamefont {van Schilfgaarde}},\ }\href {\doibase
  https://doi.org/10.1016/j.cpc.2019.107065} {\bibfield  {journal} {\bibinfo
  {journal} {Computer Physics Communications}\ ,\ \bibinfo {pages} {107065}}
  (\bibinfo {year} {2019})}\BibitemShut {NoStop}%
\bibitem [{\citenamefont {Methfessel}\ \emph {et~al.}(2000)\citenamefont
  {Methfessel}, \citenamefont {van Schilfgaarde},\ and\ \citenamefont
  {Casali}}]{Methfessel}%
  \BibitemOpen
  \bibfield  {author} {\bibinfo {author} {\bibfnamefont {M.}~\bibnamefont
  {Methfessel}}, \bibinfo {author} {\bibfnamefont {M.}~\bibnamefont {van
  Schilfgaarde}}, \ and\ \bibinfo {author} {\bibfnamefont {R.~A.}\ \bibnamefont
  {Casali}},\ }in\ \href@noop {} {\emph {\bibinfo {booktitle} {Electronic
  Structure and Physical Properties of Solids. The Use of the LMTO Method}}},\
  \bibinfo {series} {Lecture Notes in Physics}, Vol.\ \bibinfo {volume} {535},\
  \bibinfo {editor} {edited by\ \bibinfo {editor} {\bibfnamefont
  {H.}~\bibnamefont {Dreyss{\'e}}}}\ (\bibinfo  {publisher} {Berlin Springer
  Verlag},\ \bibinfo {year} {2000})\ p.\ \bibinfo {pages} {114}\BibitemShut
  {NoStop}%
\bibitem [{\citenamefont {Kotani}\ and\ \citenamefont {van
  Schilfgaarde}(2010)}]{Kotani10}%
  \BibitemOpen
  \bibfield  {author} {\bibinfo {author} {\bibfnamefont {T.}~\bibnamefont
  {Kotani}}\ and\ \bibinfo {author} {\bibfnamefont {M.}~\bibnamefont {van
  Schilfgaarde}},\ }\href {\doibase 10.1103/PhysRevB.81.125117} {\bibfield
  {journal} {\bibinfo  {journal} {Phys. Rev. B}\ }\textbf {\bibinfo {volume}
  {81}},\ \bibinfo {pages} {125117} (\bibinfo {year} {2010})}\BibitemShut
  {NoStop}%
\bibitem [{\citenamefont {Giannozzi}\ \emph {et~al.}(2009)\citenamefont
  {Giannozzi}, \citenamefont {Baroni}, \citenamefont {Bonini}, \citenamefont
  {Calandra}, \citenamefont {Car}, \citenamefont {Cavazzoni}, \citenamefont
  {Ceresoli}, \citenamefont {Chiarotti}, \citenamefont {Cococcioni},
  \citenamefont {Dabo}, \citenamefont {Corso}, \citenamefont {de~Gironcoli},
  \citenamefont {Fabris}, \citenamefont {Fratesi}, \citenamefont {Gebauer},
  \citenamefont {Gerstmann}, \citenamefont {Gougoussis}, \citenamefont
  {Kokalj}, \citenamefont {Lazzeri}, \citenamefont {Martin-Samos},
  \citenamefont {Marzari}, \citenamefont {Mauri}, \citenamefont {Mazzarello},
  \citenamefont {Paolini}, \citenamefont {Pasquarello}, \citenamefont
  {Paulatto}, \citenamefont {Sbraccia}, \citenamefont {Scandolo}, \citenamefont
  {Sclauzero}, \citenamefont {Seitsonen}, \citenamefont {Smogunov},
  \citenamefont {Umari},\ and\ \citenamefont {Wentzcovitch}}]{Giannozzi_2009}%
  \BibitemOpen
  \bibfield  {author} {\bibinfo {author} {\bibfnamefont {P.}~\bibnamefont
  {Giannozzi}}, \bibinfo {author} {\bibfnamefont {S.}~\bibnamefont {Baroni}},
  \bibinfo {author} {\bibfnamefont {N.}~\bibnamefont {Bonini}}, \bibinfo
  {author} {\bibfnamefont {M.}~\bibnamefont {Calandra}}, \bibinfo {author}
  {\bibfnamefont {R.}~\bibnamefont {Car}}, \bibinfo {author} {\bibfnamefont
  {C.}~\bibnamefont {Cavazzoni}}, \bibinfo {author} {\bibfnamefont
  {D.}~\bibnamefont {Ceresoli}}, \bibinfo {author} {\bibfnamefont {G.~L.}\
  \bibnamefont {Chiarotti}}, \bibinfo {author} {\bibfnamefont {M.}~\bibnamefont
  {Cococcioni}}, \bibinfo {author} {\bibfnamefont {I.}~\bibnamefont {Dabo}},
  \bibinfo {author} {\bibfnamefont {A.~D.}\ \bibnamefont {Corso}}, \bibinfo
  {author} {\bibfnamefont {S.}~\bibnamefont {de~Gironcoli}}, \bibinfo {author}
  {\bibfnamefont {S.}~\bibnamefont {Fabris}}, \bibinfo {author} {\bibfnamefont
  {G.}~\bibnamefont {Fratesi}}, \bibinfo {author} {\bibfnamefont
  {R.}~\bibnamefont {Gebauer}}, \bibinfo {author} {\bibfnamefont
  {U.}~\bibnamefont {Gerstmann}}, \bibinfo {author} {\bibfnamefont
  {C.}~\bibnamefont {Gougoussis}}, \bibinfo {author} {\bibfnamefont
  {A.}~\bibnamefont {Kokalj}}, \bibinfo {author} {\bibfnamefont
  {M.}~\bibnamefont {Lazzeri}}, \bibinfo {author} {\bibfnamefont
  {L.}~\bibnamefont {Martin-Samos}}, \bibinfo {author} {\bibfnamefont
  {N.}~\bibnamefont {Marzari}}, \bibinfo {author} {\bibfnamefont
  {F.}~\bibnamefont {Mauri}}, \bibinfo {author} {\bibfnamefont
  {R.}~\bibnamefont {Mazzarello}}, \bibinfo {author} {\bibfnamefont
  {S.}~\bibnamefont {Paolini}}, \bibinfo {author} {\bibfnamefont
  {A.}~\bibnamefont {Pasquarello}}, \bibinfo {author} {\bibfnamefont
  {L.}~\bibnamefont {Paulatto}}, \bibinfo {author} {\bibfnamefont
  {C.}~\bibnamefont {Sbraccia}}, \bibinfo {author} {\bibfnamefont
  {S.}~\bibnamefont {Scandolo}}, \bibinfo {author} {\bibfnamefont
  {G.}~\bibnamefont {Sclauzero}}, \bibinfo {author} {\bibfnamefont {A.~P.}\
  \bibnamefont {Seitsonen}}, \bibinfo {author} {\bibfnamefont {A.}~\bibnamefont
  {Smogunov}}, \bibinfo {author} {\bibfnamefont {P.}~\bibnamefont {Umari}}, \
  and\ \bibinfo {author} {\bibfnamefont {R.~M.}\ \bibnamefont {Wentzcovitch}},\
  }\href {\doibase 10.1088/0953-8984/21/39/395502} {\bibfield  {journal}
  {\bibinfo  {journal} {Journal of Physics: Condensed Matter}\ }\textbf
  {\bibinfo {volume} {21}},\ \bibinfo {pages} {395502} (\bibinfo {year}
  {2009})}\BibitemShut {NoStop}%
\bibitem [{\citenamefont {Perdew}\ \emph {et~al.}(1996)\citenamefont {Perdew},
  \citenamefont {Burke},\ and\ \citenamefont {Ernzerhof}}]{PBE}%
  \BibitemOpen
  \bibfield  {author} {\bibinfo {author} {\bibfnamefont {J.~P.}\ \bibnamefont
  {Perdew}}, \bibinfo {author} {\bibfnamefont {K.}~\bibnamefont {Burke}}, \
  and\ \bibinfo {author} {\bibfnamefont {M.}~\bibnamefont {Ernzerhof}},\ }\href
  {\doibase 10.1103/PhysRevLett.77.3865} {\bibfield  {journal} {\bibinfo
  {journal} {Phys. Rev. Lett.}\ }\textbf {\bibinfo {volume} {77}},\ \bibinfo
  {pages} {3865} (\bibinfo {year} {1996})}\BibitemShut {NoStop}%
\bibitem [{\citenamefont {Hedin}(1965)}]{Hedin65}%
  \BibitemOpen
  \bibfield  {author} {\bibinfo {author} {\bibfnamefont {L.}~\bibnamefont
  {Hedin}},\ }\href {\doibase 10.1103/PhysRev.139.A796} {\bibfield  {journal}
  {\bibinfo  {journal} {Phys. Rev.}\ }\textbf {\bibinfo {volume} {139}},\
  \bibinfo {pages} {A796} (\bibinfo {year} {1965})}\BibitemShut {NoStop}%
\bibitem [{\citenamefont {Hedin}\ and\ \citenamefont
  {Lundqvist}(1969)}]{Hedin69}%
  \BibitemOpen
  \bibfield  {author} {\bibinfo {author} {\bibfnamefont {L.}~\bibnamefont
  {Hedin}}\ and\ \bibinfo {author} {\bibfnamefont {S.}~\bibnamefont
  {Lundqvist}},\ }in\ \href@noop {} {\emph {\bibinfo {booktitle} {Solid State
  Physics, Advanced in Research and Applications}}},\ Vol.~\bibinfo {volume}
  {23},\ \bibinfo {editor} {edited by\ \bibinfo {editor} {\bibfnamefont
  {F.}~\bibnamefont {Seitz}}, \bibinfo {editor} {\bibfnamefont
  {D.}~\bibnamefont {Turnbull}}, \ and\ \bibinfo {editor} {\bibfnamefont
  {H.}~\bibnamefont {Ehrenreich}}}\ (\bibinfo  {publisher} {Academic Press},\
  \bibinfo {address} {New York},\ \bibinfo {year} {1969})\ pp.\ \bibinfo
  {pages} {1--181}\BibitemShut {NoStop}%
\bibitem [{\citenamefont {Kotani}\ \emph {et~al.}(2007)\citenamefont {Kotani},
  \citenamefont {van Schilfgaarde},\ and\ \citenamefont {Faleev}}]{Kotani07}%
  \BibitemOpen
  \bibfield  {author} {\bibinfo {author} {\bibfnamefont {T.}~\bibnamefont
  {Kotani}}, \bibinfo {author} {\bibfnamefont {M.}~\bibnamefont {van
  Schilfgaarde}}, \ and\ \bibinfo {author} {\bibfnamefont {S.~V.}\ \bibnamefont
  {Faleev}},\ }\href {\doibase 10.1103/PhysRevB.76.165106} {\bibfield
  {journal} {\bibinfo  {journal} {Phys.Rev. B}\ }\textbf {\bibinfo {volume}
  {76}},\ \bibinfo {eid} {165106} (\bibinfo {year} {2007})}\BibitemShut
  {NoStop}%
\bibitem [{\citenamefont {van Schilfgaarde}\ \emph {et~al.}(2006)\citenamefont
  {van Schilfgaarde}, \citenamefont {Kotani},\ and\ \citenamefont
  {Faleev}}]{vanSchilfgaarde06}%
  \BibitemOpen
  \bibfield  {author} {\bibinfo {author} {\bibfnamefont {M.}~\bibnamefont {van
  Schilfgaarde}}, \bibinfo {author} {\bibfnamefont {T.}~\bibnamefont {Kotani}},
  \ and\ \bibinfo {author} {\bibfnamefont {S.~V.}\ \bibnamefont {Faleev}},\
  }\href {\doibase 10.1103/PhysRevB.74.245125} {\bibfield  {journal} {\bibinfo
  {journal} {Phys.Rev. B}\ }\textbf {\bibinfo {volume} {74}},\ \bibinfo {eid}
  {245125} (\bibinfo {year} {2006})}\BibitemShut {NoStop}%
\bibitem [{\citenamefont {Marezio}\ and\ \citenamefont
  {Remeika}(1965)}]{MarezioRemeika}%
  \BibitemOpen
  \bibfield  {author} {\bibinfo {author} {\bibfnamefont {M.}~\bibnamefont
  {Marezio}}\ and\ \bibinfo {author} {\bibfnamefont {J.}~\bibnamefont
  {Remeika}},\ }\href {\doibase https://doi.org/10.1016/0022-3697(65)90108-3}
  {\bibfield  {journal} {\bibinfo  {journal} {Journal of Physics and Chemistry
  of Solids}\ }\textbf {\bibinfo {volume} {26}},\ \bibinfo {pages} {1277 }
  (\bibinfo {year} {1965})}\BibitemShut {NoStop}%
\bibitem [{\citenamefont {Lei}\ \emph {et~al.}(2010)\citenamefont {Lei},
  \citenamefont {Irifune}, \citenamefont {Shinmei}, \citenamefont {Ohfuji},\
  and\ \citenamefont {Fang}}]{Lei10}%
  \BibitemOpen
  \bibfield  {author} {\bibinfo {author} {\bibfnamefont {L.}~\bibnamefont
  {Lei}}, \bibinfo {author} {\bibfnamefont {T.}~\bibnamefont {Irifune}},
  \bibinfo {author} {\bibfnamefont {T.}~\bibnamefont {Shinmei}}, \bibinfo
  {author} {\bibfnamefont {H.}~\bibnamefont {Ohfuji}}, \ and\ \bibinfo {author}
  {\bibfnamefont {L.}~\bibnamefont {Fang}},\ }\href {\doibase
  10.1063/1.3487976} {\bibfield  {journal} {\bibinfo  {journal} {Journal of
  Applied Physics}\ }\textbf {\bibinfo {volume} {108}},\ \bibinfo {pages}
  {083531} (\bibinfo {year} {2010})}\BibitemShut {NoStop}%
\bibitem [{\citenamefont {Lei}\ \emph {et~al.}(2013)\citenamefont {Lei},
  \citenamefont {Ohfuji}, \citenamefont {Qin}, \citenamefont {Zhang},
  \citenamefont {Wang},\ and\ \citenamefont {Irifune}}]{Lei13}%
  \BibitemOpen
  \bibfield  {author} {\bibinfo {author} {\bibfnamefont {L.}~\bibnamefont
  {Lei}}, \bibinfo {author} {\bibfnamefont {H.}~\bibnamefont {Ohfuji}},
  \bibinfo {author} {\bibfnamefont {J.}~\bibnamefont {Qin}}, \bibinfo {author}
  {\bibfnamefont {X.}~\bibnamefont {Zhang}}, \bibinfo {author} {\bibfnamefont
  {F.}~\bibnamefont {Wang}}, \ and\ \bibinfo {author} {\bibfnamefont
  {T.}~\bibnamefont {Irifune}},\ }\href {\doibase
  https://doi.org/10.1016/j.ssc.2013.03.030} {\bibfield  {journal} {\bibinfo
  {journal} {Solid State Communications}\ }\textbf {\bibinfo {volume} {164}},\
  \bibinfo {pages} {6 } (\bibinfo {year} {2013})}\BibitemShut {NoStop}%
\bibitem [{\citenamefont {Koster}\ \emph {et~al.}(1963)\citenamefont {Koster},
  \citenamefont {Dimmock}, \citenamefont {Wheeler},\ and\ \citenamefont
  {Statz}}]{Koster}%
  \BibitemOpen
  \bibfield  {author} {\bibinfo {author} {\bibfnamefont {G.~F.}\ \bibnamefont
  {Koster}}, \bibinfo {author} {\bibfnamefont {J.~O.}\ \bibnamefont {Dimmock}},
  \bibinfo {author} {\bibfnamefont {R.~G.}\ \bibnamefont {Wheeler}}, \ and\
  \bibinfo {author} {\bibfnamefont {H.}~\bibnamefont {Statz}},\ }\href@noop {}
  {\emph {\bibinfo {title} {{Properties of the Thirty-Two Point Groups}}}}\
  (\bibinfo  {publisher} {MIT Press},\ \bibinfo {address} {Cambridge, MA},\
  \bibinfo {year} {1963})\BibitemShut {NoStop}%
\bibitem [{\citenamefont {Lambrecht}\ \emph {et~al.}(2002)\citenamefont
  {Lambrecht}, \citenamefont {Rodina}, \citenamefont {Limpijumnong},
  \citenamefont {Segall},\ and\ \citenamefont {Meyer}}]{Lambrecht02}%
  \BibitemOpen
  \bibfield  {author} {\bibinfo {author} {\bibfnamefont {W.~R.~L.}\
  \bibnamefont {Lambrecht}}, \bibinfo {author} {\bibfnamefont {A.~V.}\
  \bibnamefont {Rodina}}, \bibinfo {author} {\bibfnamefont {S.}~\bibnamefont
  {Limpijumnong}}, \bibinfo {author} {\bibfnamefont {B.}~\bibnamefont
  {Segall}}, \ and\ \bibinfo {author} {\bibfnamefont {B.~K.}\ \bibnamefont
  {Meyer}},\ }\href {\doibase 10.1103/PhysRevB.65.075207} {\bibfield  {journal}
  {\bibinfo  {journal} {Phys. Rev. B}\ }\textbf {\bibinfo {volume} {65}},\
  \bibinfo {pages} {075207} (\bibinfo {year} {2002})}\BibitemShut {NoStop}%
\bibitem [{\citenamefont {Tumėnas}\ \emph {et~al.}(2017)\citenamefont
  {Tumėnas}, \citenamefont {Mackonis}, \citenamefont {Nedzinskas},
  \citenamefont {Trinkler}, \citenamefont {Berzina}, \citenamefont {Korsaks},
  \citenamefont {Chang},\ and\ \citenamefont {Chou}}]{Tumenasi17}%
  \BibitemOpen
  \bibfield  {author} {\bibinfo {author} {\bibfnamefont {S.}~\bibnamefont
  {Tumėnas}}, \bibinfo {author} {\bibfnamefont {P.}~\bibnamefont {Mackonis}},
  \bibinfo {author} {\bibfnamefont {R.}~\bibnamefont {Nedzinskas}}, \bibinfo
  {author} {\bibfnamefont {L.}~\bibnamefont {Trinkler}}, \bibinfo {author}
  {\bibfnamefont {B.}~\bibnamefont {Berzina}}, \bibinfo {author} {\bibfnamefont
  {V.}~\bibnamefont {Korsaks}}, \bibinfo {author} {\bibfnamefont
  {L.}~\bibnamefont {Chang}}, \ and\ \bibinfo {author} {\bibfnamefont
  {M.}~\bibnamefont {Chou}},\ }\href {\doibase
  https://doi.org/10.1016/j.apsusc.2017.01.098} {\bibfield  {journal} {\bibinfo
   {journal} {Applied Surface Science}\ }\textbf {\bibinfo {volume} {421}},\
  \bibinfo {pages} {837 } (\bibinfo {year} {2017})}\BibitemShut {NoStop}%
\bibitem [{\citenamefont {Levine}\ and\ \citenamefont
  {Allan}(1989)}]{LevineAllan89}%
  \BibitemOpen
  \bibfield  {author} {\bibinfo {author} {\bibfnamefont {Z.~H.}\ \bibnamefont
  {Levine}}\ and\ \bibinfo {author} {\bibfnamefont {D.~C.}\ \bibnamefont
  {Allan}},\ }\href {\doibase 10.1103/PhysRevLett.63.1719} {\bibfield
  {journal} {\bibinfo  {journal} {Phys. Rev. Lett.}\ }\textbf {\bibinfo
  {volume} {63}},\ \bibinfo {pages} {1719} (\bibinfo {year}
  {1989})}\BibitemShut {NoStop}%
\bibitem [{\citenamefont {Cunningham}\ \emph {et~al.}(2018)\citenamefont
  {Cunningham}, \citenamefont {Gr\"uning}, \citenamefont {Azarhoosh},
  \citenamefont {Pashov},\ and\ \citenamefont {van
  Schilfgaarde}}]{Cunningham18}%
  \BibitemOpen
  \bibfield  {author} {\bibinfo {author} {\bibfnamefont {B.}~\bibnamefont
  {Cunningham}}, \bibinfo {author} {\bibfnamefont {M.}~\bibnamefont
  {Gr\"uning}}, \bibinfo {author} {\bibfnamefont {P.}~\bibnamefont
  {Azarhoosh}}, \bibinfo {author} {\bibfnamefont {D.}~\bibnamefont {Pashov}}, \
  and\ \bibinfo {author} {\bibfnamefont {M.}~\bibnamefont {van Schilfgaarde}},\
  }\href {\doibase 10.1103/PhysRevMaterials.2.034603} {\bibfield  {journal}
  {\bibinfo  {journal} {Phys. Rev. Materials}\ }\textbf {\bibinfo {volume}
  {2}},\ \bibinfo {pages} {034603} (\bibinfo {year} {2018})}\BibitemShut
  {NoStop}%
\bibitem [{\citenamefont {Bhandari}\ \emph {et~al.}(2018)\citenamefont
  {Bhandari}, \citenamefont {van Schilfgaarde}, \citenamefont {Kotani},\ and\
  \citenamefont {Lambrecht}}]{Churna18}%
  \BibitemOpen
  \bibfield  {author} {\bibinfo {author} {\bibfnamefont {C.}~\bibnamefont
  {Bhandari}}, \bibinfo {author} {\bibfnamefont {M.}~\bibnamefont {van
  Schilfgaarde}}, \bibinfo {author} {\bibfnamefont {T.}~\bibnamefont {Kotani}},
  \ and\ \bibinfo {author} {\bibfnamefont {W.~R.~L.}\ \bibnamefont
  {Lambrecht}},\ }\href {\doibase 10.1103/PhysRevMaterials.2.013807} {\bibfield
   {journal} {\bibinfo  {journal} {Phys. Rev. Materials}\ }\textbf {\bibinfo
  {volume} {2}},\ \bibinfo {pages} {013807} (\bibinfo {year}
  {2018})}\BibitemShut {NoStop}%
\bibitem [{\citenamefont {Lambrecht}\ \emph {et~al.}(2017)\citenamefont
  {Lambrecht}, \citenamefont {Bhandari},\ and\ \citenamefont {van
  Schilfgaarde}}]{Lambrecht17}%
  \BibitemOpen
  \bibfield  {author} {\bibinfo {author} {\bibfnamefont {W.~R.~L.}\
  \bibnamefont {Lambrecht}}, \bibinfo {author} {\bibfnamefont {C.}~\bibnamefont
  {Bhandari}}, \ and\ \bibinfo {author} {\bibfnamefont {M.}~\bibnamefont {van
  Schilfgaarde}},\ }\href {\doibase 10.1103/PhysRevMaterials.1.043802}
  {\bibfield  {journal} {\bibinfo  {journal} {Phys. Rev. Materials}\ }\textbf
  {\bibinfo {volume} {1}},\ \bibinfo {pages} {043802} (\bibinfo {year}
  {2017})}\BibitemShut {NoStop}%
\bibitem [{\citenamefont {Verdi}\ and\ \citenamefont
  {Giustino}(2015)}]{Verdi15}%
  \BibitemOpen
  \bibfield  {author} {\bibinfo {author} {\bibfnamefont {C.}~\bibnamefont
  {Verdi}}\ and\ \bibinfo {author} {\bibfnamefont {F.}~\bibnamefont
  {Giustino}},\ }\href {\doibase 10.1103/PhysRevLett.115.176401} {\bibfield
  {journal} {\bibinfo  {journal} {Phys. Rev. Lett.}\ }\textbf {\bibinfo
  {volume} {115}},\ \bibinfo {pages} {176401} (\bibinfo {year}
  {2015})}\BibitemShut {NoStop}%
\end{thebibliography}%
\end{document}